\documentclass[aps,twocolumn, superscriptaddress,groupedaddress,nofootinbib,10pt,preprintnumbers]{revtex4-1}
\usepackage{graphicx}  
\usepackage{dcolumn}   
\usepackage{bm,bbm}        
\usepackage{amssymb}   
\usepackage{amsmath}
\usepackage{amsfonts}
\usepackage{array}
\usepackage{xcolor}

\usepackage{tikz}
\usetikzlibrary{shapes.geometric, arrows}
\usepackage{lipsum}
\usepackage{float}
\usepackage{multirow}
\usepackage{hhline}
\usepackage{tabularx}
\usepackage{subfigure}
\usepackage{graphics}
\usepackage{amsthm}
\usepackage{mathtools}
\usepackage{amsmath,amsfonts,amssymb}       
\usepackage{appendix}
\usepackage{graphicx}
\usepackage{color}

\theoremstyle{definition}

\theoremstyle{definition}

\usepackage{algpseudocode}
\usepackage{algorithm}

\newcommand{\cO}{\mathcal{O}}
\newcommand{\bE}{\mathbb{E}}
\newcommand{\bR}{\mathbb{R}}

\newcommand{\Ng}{N_\text{geom}}
\newcommand{\Nb}{N_\text{batch}}
\newcommand{\Nc}{N_\text{crit}}

\begin{document}


\title{Statistical Predictions in String Theory and Deep Generative Models}
\author{James Halverson}
\author{Cody Long}
\affiliation{Department of Physics, Northeastern University \\ Boston, MA 02115-5000 USA} 

\date{\today}

\begin{abstract}
Generative models in deep learning allow
for sampling probability distributions that
approximate data distributions. We propose
using generative models for making approximate
statistical predictions in the string theory
landscape. 
For vacua admitting
a Lagrangian description this can be thought of as learning random tensor approximations
of couplings. As a concrete proof-of-principle, we demonstrate in
a large ensemble of Calabi-Yau manifolds
that K\" ahler metrics evaluated at points
in K\" ahler moduli space are
well-approximated by ensembles of matrices
produced by a deep convolutional Wasserstein
GAN.
Accurate approximations of the K\" ahler
metric eigenspectra are achieved
with far fewer than $h^{11}$
Gaussian draws. Accurate extrapolation to
values of $h^{11}$ outside the training
set are achieved via a conditional GAN.
Together, these results implicitly suggest the existence of strong correlations
in the data, as might be expected if Reid's fantasy is correct.
\end{abstract}

\pacs{}
\maketitle

\section{Introduction}
String theory is a leading candidate for
unifying quantum gravity with
particle physics and cosmology. It has 
a large landscape of vacua, due not only
to the plethora of fluxes \cite{Bousso:2000xa,Denef:2004ze,Denef:2004cf,Taylor:2015xtz} that may exist in its extra dimensions,
but also  the large number of extra-dimensional geometries 
\cite{Kreuzer:2000xy,Halverson:2017ffz,Taylor:2017yqr,Altman:2018zlc} themselves; the known size of both has
grown significantly in recent years. The landscape
gives rise to rich
cosmological dynamics
and diverse low energy compactifications that
may exhibit
features of the Standard Models of
particle physics and cosmology, as well
as
observable remnants of the ultraviolet theory.
If string theory is true, fundamental
physics is a complex system.

The diversity of possibilities has a simple implication: predictions in string theory are statistical \cite{Douglas:2003um}. Given
the distribution $P(i)$ on vacua, one
would like to compute expectation values
of observables $\cO$  
\begin{equation}
\bE_{i\sim P(i)}[\mathcal{O}]= \sum_{i\in S_\text{vac}} D(i) A(i) \,\,\cO (i),
\end{equation}
where we have written $P(i)=D(i)A(i)$ in
a factorized form involving a dynamical
factor $D(i)$ and an anthropic factor
$A(i)$. These factors give important corrections from a naive uniform
distribution. Unfortunately, neither the full
set of vacua $S_\text{vac}$ nor the factors
$D(i)$ or $A(i)$ are currently known in full,
and significant theoretical work is required
to determine them. However, when drawing
from a uniform distribution, the largest known sets of flux
vacua \cite{Taylor:2015xtz} and geometries
\cite{Halverson:2017ffz,Taylor:2017yqr} both
suggest that large numbers of gauge sectors
and axion-like particles are the rule, not
the exception; both have significant
cosmological implications, see, e.g., \cite{Halverson:2016nfq,Halverson:2019kna,Halverson:2019cmy}. A number of proposals exist
for the dynamical factor, including global measures \cite{Garriga:2005av,DeSimone:2008if,Carifio:2017nyb}, local measures \cite{Bousso:2006ge,Bousso:2006xc,Bousso:2008hz,Freivogel:2011eg}, and computational complexity based
measures \cite{Denef:2017cxt,Khoury:2019yoo,Khoury:2019ajl}. 
It is even more difficult to compute
$A(i)$, butthere is significant evidence that
it depends on the cosmological constant \cite{Weinberg:1987dv,Bousso:2000xa}.

Though the full extent to which computational
complexity affects the dynamical factor is
not known, it certainly affects practical
efforts to study the landscape. That is,
difficulties arises not only from
the large number of vacua, but also 
the computational complexity of physical
questions related to them. For instance,
finding small cosmological constants
in the Bousso-Polchinski model is
NP-complete \cite{Denef:2006ad},
solving decision problems in the landscape runs up against
Diophantine undecidability \cite{Cvetic:2010ky},
and both constructing and minimizing the
scalar potential requires \cite{Halverson:2018cio} solving instances of NP-hard and co-NP-hard problems. In some cases the structure of the theory may allow for the avoidance of worst-case
complexity, for instance in ED3-instanton problems \cite{Halverson:2019vmd}, but identifying
such instances can itself be challenging. 

Alternatively,
approximations can allow for the avoidance of complexity. For instance, so-called fully-polynomial
time approximation schemes are algorithms for solving a problem with error bounded by 
$\epsilon$, such that the runtime is polynomial in the input size and $1/\epsilon$,
even if the exact version of the problem is NP-hard.

\bigskip

Complexity provides a significant obstacle to
making statistical predictions in string compactifications.
As such, it is natural
to wonder whether one could get by with making
approximate predictions,
using appropriate approximations
\begin{equation}
\hat S_\text{vac} \simeq S_\text{vac}, \qquad
\hat D(i) \simeq D(i), \qquad
\hat A(i) \simeq A(i). 
\end{equation}
Rather than computing expectation values
of observables
in an exact distribution on string vacua,
one could attempt to make predictions 
in an approximate distribution. Put differently, if 
exact calculations in string theory are too slow, could fast-but-accurate
simulation suffice?

Determining approximations to distributions of
data $P_d$ is the subject of so-called generative
models in deep learning.\footnote{Applications of deep
learning to string theory have been of recent interest.
See \cite{He:2017aed,Krefl:2017yox,Ruehle:2017mzq,Carifio:2017bov} for original works, \cite{Liu:2017dzi,
*Carifio:2017nyb,
*Hashimoto:2018ftp,
*Wang:2018rkk,
*Jinno:2018dek,
*Bull:2018uow,
*Constantin:2018hvl,
*Klaewer:2018sfl,
*Rudelius:2018yqi,
*Mutter:2018sra,
*Altman:2018zlc,
*He:2018jtw,
*Cole:2018emh,
*Jejjala:2019kio,
*Bull:2019cij,
*Hashimoto:2019bih,
*Halverson:2019tkf,
*He:2019vsj,
*He:2019nzx,
*Cole:2019enn,
*Ashmore:2019wzb,
*Parr:2019bta,
*Alessandretti:2019jbs}
for further progress with
a variety of techniques,
and \cite{RUEHLE2019} for a review.} In generative models,
a random variable $z$ drawn from $P(z)$, written $z \sim P(z)$, is passed
through a parameterized function 
\begin{equation}
F_\theta: \mathcal{Z}\to
\mathcal{D}\, ,
\end{equation} generating a sample $F_\theta(z)$ of an
implicit distribution $P_{\theta}$ that
depends on the parameters $\theta$ in
$F_\theta$; often, $F_\theta$ is a deep
neural network. There are many classes 
of generative models, corresponding to
different functional forms for $F_\theta$ and different algorithms for
optimizing its parameters.
The optimization procedure leads to increasingly good approximations
$P_\theta\simeq P_d$, as measured by an
appropriate distance measure such as the 
Kullback-Leibler divergence or the Wasserstein
distance.

\bigskip

In this paper we explore utilizing generative models to
make statistical predictions in string theory. 

This is a rather
general idea, but it is very concrete for vacua admitting
a Lagrangian description, where modeling statistical predictions
involves learning random tensor approximations of the 
couplings in the Lagrangian. For matrix couplings, such as
mass matrices or metrics appearing in kinetic terms, this
amounts to directly learning a random matrix ensemble that simulates
string data, rather than attempting to guess one a priori.

As a direct application, we use a class of generative models
known as generative adversarial networks (GANs) \cite{gan} to learn a random
matrix approximation to K\" ahler metrics evaluated at points on the K\" ahler moduli
space of Calabi-Yau manifolds.\footnote{See \cite{Erbin:2018csv}
for a study of the use of GANs in generating EFTs.}
In each case, a generator neural network $G_\theta$ is
trained on K\" ahler metrics obtained in Calabi-Yau compactification by optimization of the parameters
$\theta$. For any epoch in the training, $G_\theta$
can be used to generate simulated K\" ahler metrics
that model real ones increasingly well as training
proceeds, as measured for instance by the 
Wasserstein distance on the log eigenspectra
of the real and simulated K\" ahler metrics.

For those less familiar with generative models, we would
like to highlight the important role of the noise $z\sim P(z)$, which has dimension $n_z$, used in simulation. The training process involves optimizing the network $G_\theta$ so that noise sent through the network models data, where in general
one expects that the value of $n_z$ affects the quality 
of the model, i.e. there is some minimal dimension of input necessary to model the data. In models where GANs are
trained at fixed values of $h^{11}$, we find that performance
is relatively insensitive to $n_z$, provided $n_z \gtrsim 5$,
which itself is much below $h^{11}$. This implicit
suggests the existence of correlations in the data.

Since many string vacua arise at large\footnote{Throughout,
when we are vague about $N$ in string theory it is a proxy for the number of degrees of freedom, fluxes, cycles, etc.}
 $N$, where calculations are often intractable, it would be
useful to have a fast-but-accurate simulator of string
data in that regime. This requires extrapolation outside of
the training sample, which a priori is difficult unless
significant structure exists in the data that allows for
extrapolation. We demonstrate that a conditional GAN
gives rise to better-than-expected extrapolation to
larger values of $h^{11}$ for K\" ahler metrics.

\medskip
This paper is organized as follows. In
Section \ref{sec:generative_intro} we review
generative models and introduce the original GAN
and the Wasserstein GAN. In Section \ref{sec:random_tensor_approx} we introduce the use of
generative models for learning random tensor approximations
of string effective Lagrangians, exemplifying the idea
for K\" ahler metrics at fixed $h^{11}$ in Section \ref{sec:fix_h11} and extrapolating to $h^{11}$ values outside
the training set in Section \ref{sec:interp_extrap}.
In Section \ref{sec:discussion} we review the main results. That section is the primary location where results
are discussed, in an effort to
separate the implications of the results from
somewhat technical deep learning details utilized in their
derivation.

\section{Generative Models}
\label{sec:generative_intro}

We have already introduced the essential idea behind generative models:
to learn how to generate samples of a distribution that closely approximates
a data distribution, i.e., to produce reliable (and fast) simulations.

A myriad of generative models exist. Common models not utilized in this work include variational autoencoders \cite{kingma2013autoencoding}
and normalizing flows \cite{rezende2015variational}. The former provide a modification of
an autoencoder architecture such that the second half of a network may be used to generate data
given draws from a multivariate Gaussian, while the latter focuses on utilizing an invertible architecture, which in turn  allows for the evaluation of sample probabilities via inversion. Potential uses of these techniques in string theory will be discussed in Section \ref{sec:discussion}.

\medskip

The generative models that we utilize in this work
are generative adversarial networks \cite{gan} (GANs). GANs pit a generator
network against an adversary network, often referred to as a 
discriminator or critic, where the training goal of the generator
is to produce fake samples from noise that fool the adversary,
while the latter discriminator is trained to determine whether the samples it sees
are real or were produced by the generator. 
We will utilize a variety of GANs,
which differ according to their loss functions \cite{gan,wgan} and network
architecture \cite{dcgan}. For the purposes of interpolation
and extrapolation, we will also pass conditions to the GAN
\cite{cgan}, where in our case the condition will be a value of $h^{11}$
for which to simulate a K\" ahler metric. In the end, we will find that a Wasserstein GAN
with deep convolutional architecture outperforms the others.\footnote{A comparative study \cite{equalgans} of the performance of different GANs 
suggests that fine-tuning of hyperparameters can sometimes compensate
for fundamental algorithmic differences.}

Let us review the original GAN \cite{gan} as we utilize it in this
paper. It consists of a generator and a discriminator network
\begin{align}
G_\theta&: \bR^{n_z} \to \bR^{N} \times \bR^{N} \nonumber \\
D_w&: \bR^{N} \times \bR^{N} \to [0,1]\, ,
\end{align}
parameterized by $\theta$ and $w$, respectively. Real and fake
data are labelled $1$ and $0$, respectively. We will
sometimes abbreviate $G_\theta$ and $D_w$ to $G$ and $D$. 
During training, the generator is trained on
a batch of simulated data $G(z)$ generated from a batch of noise
$z\sim P(z)$ drawn from a noise distribution. The discriminator
is trained on equal-size batches of real data $x\sim P_d(x)$ and
simulated data $G(z)$. From the batches, the parameters are updated according to the
loss functions
\begin{align}
L_D^\text{GAN} &= -\bE_{x\sim P_d(x)}[log(D(x)] + \bE_{z\sim P(z)}[log(1-D(G(z)))] \nonumber \\
L_G^\text{GAN} & = -\bE_{z\sim P(z)}[log(D(G(z)))]\, .
\end{align}
These losses may be interpreted term by term. For instance, consider
a given $z\sim P(z)$ such that 
the discriminator thinks the generated data is real, i.e.
$D(G(z)) = 1$. In this case the simulated $G(z)$ 
does not contribute to $L_G^\text{GAN}$ but gives a large contribution
to $L_D^\text{GAN}$, which is as desired since the generator has fooled
the discriminator into thinking that $G(z)$ is real. The 
converse holds as expected if $D(G(z))=0$, the generator is penalized
since the discriminator has detected that $G(z)$ is a fake. Similarly,
real data $x\sim P_d(x)$ penalizes $D$ by a positive contribution to
$L_D^\text{GAN}$ when $D(x) < 0$ i.e. when the discriminator is not
sure that $x$ is real.

\bigskip
We also utilize Wasserstein GANs (WGANs)~\cite{wgan}, which differ from
the original GAN in important ways. There is an intuitive
understanding and a more formal one; we begin by describing the latter, which will
lead to an intuitive understanding after  an approximation.

The WGAN is built
on a solid theoretical foundation. Following \cite{wgan},
consider ways in which to measure how close the model
distribution $P_\theta$ is to the real distribution $P_d$,
as measured by $\rho(P_\theta,P_d)$; in some cases $\rho$ is a proper
distance, but in other often-used cases it is a divergence (such
as the Kullback-Leibler divergence) that is not symmetric in the
two distributions. A sequence of distributions $P_t$ with $t\in \mathbb{N}$ is said to converge (in $\rho$) to a distribution $P_\infty$ if 
$\rho(P_t,P_\infty)$ goes to zero as $t\to \infty$. Given
two distances or divergences $\rho$ and $\rho'$, if the 
set of sequences convergent under $\rho$ is a superset of 
those convergent under $\rho'$, then it is said that $\rho$ 
induces a weaker topology; that is, has better convergence properties.

Since a GAN learns an implicit
probability distribution, a natural learning question is which
distance or divergence has the weakest topology, i.e. will have the
best convergence properties to the data distribution. 
Four possibilities
are considered in \cite{wgan}, according to whether $\rho$ is the
total variation (TV) distance, the Kullback-Leibler (KL) divergence, 
the Jensen-Shannon (JS) divergence, or earth-mover (EM) distance,
which is also known as the Wasserstein distance. The main
theorem in~\cite{wgan} shows that the Wasserstein distance has the best
convergence properties, followed by JS and TV, followed by KL, 
suggesting the utilizing the Wasserstein distance in a GAN
could lead to superior training.

Unfortunately, the Wasserstein distance 
\begin{equation}
W(P_\theta,P_d) = \inf_{\gamma \in \Pi(P_d,P_\theta)} \bE_{(x,y)\sim \gamma}[\,||x-y||\,]\, ,
\end{equation}
is intractable to compute for high dimensional distributions. Here $\Pi(P_d,P_\theta)$ are all joint distributions whose marginals
are $P_d$ and $P_\theta$.
However, Kantorovich-Rubinstein duality (see, e.g., \cite{KRduality}) allows it to be rewritten as
\begin{equation}
\label{eq:wass_f_approx}
W(P_d,P_\theta) = \frac{1}{K} \,\, \sup_{||f||_L\leq K} \mathbb{E}_{x\sim P_d}[f(x)] - \mathbb{E}_{y\sim P_\theta}[f(y)]\, ,
\end{equation}
which involves the supremum over all functions $f$ to $\bR$ that are $K$-Lipschitz for some constant $K$, i.e. $|f(a)-f(b)| \leq K\times |a-b|$ for all $a,b$ on the domain, denoted $||f||_L \leq K$. One could instead consider maximizing over a family
of functions $f_w$ parameterized by $w\in W$, where compact $W$
ensures that $f_w$ is $K$-Lipschitz for some $K$. For instance
$f_w$ could be a neural network with weights clamped to a compact
space.

With this introduction, we can now reintroduce the generator $G_\theta$.
Let $P_d$ again be the data distribution, and $P_\theta$
be the implicit distribution of $G_\theta(z)$ with noise $z\in p(z)$. Let
what was called $f_w$ be the discriminator $D_w$. Then under
suitable assumptions \cite{wgan}
\begin{align}
\label{eq:wgan_generator_update}
\nabla_\theta W(P_d,P_\theta) &= -\nabla_\theta\,\, \bE_{z\sim p(z)}[D_w(G_\theta(z))] \nonumber \\ 
&= - \bE_{z\sim p(z)}[\nabla_\theta D_w(G_\theta(z))]\, ,
\end{align}
which is readibly computable. Here it is implicit
that $D_w$ has been trained to
be the function $f$ in \eqref{eq:wass_f_approx}
that appears in the approximation of $W(P_\theta,P_d)$. We then have
a simple gradient for the generator update that approximates (if $D_w$
is perfectly trained) the
Wasserstein gradient. This update, together with
the discriminator update \eqref{eq:wass_f_approx},
form the basis of the Wasserstein GAN \cite{wgan}.

We now see a significant qualitative departure from
many GANs. In training typical GAN it is often 
possible to over-train the discriminator, leading to
to poor gradient updates for the generator. 
For the generator gradient update in~\eqref{eq:wgan_generator_update} for
a WGAN, 
note instead the RHS only approximates the
gradient of the Wasserstein distance (a useful
gradient for training) when $D_w$ itself is well-trained. That is, the generator receives
useful updates when the discriminator is strong.
For this reason, the WGAN discriminator is often instead
called a critic; its goal is not to be
an adversary to the generator with which it competes, but instead an expert data connoisseur  that helps the generator improve its behavior.

After this theoretical development, an
intuitive understanding of the WGAN may be useful.
Keeping in mind that the critic $D_w$ is trained
to play the role of the function $f$ in~\eqref{eq:wass_f_approx}, we see that it is training to obtain
maximal separation between
the real data and the fake data, i.e. maximizing
the difference
\begin{equation}
\mathbb{E}_{x\sim P_d}[D_w(x)] - \mathbb{E}_{y\sim P_\theta}[D_w(y)] \in \bR,
\end{equation}
or alternatively minimizing its negative.
The generator loss in \eqref{eq:wgan_generator_update}
simply attempts to push $D_w(y)=D_w(G_\theta(z))$ the 
other direction, leading to a competition.

\bigskip
Finally, we introduce the conditional GAN \cite{cgan} (cGAN).
The idea behind the cGAN is rather simple: in some cases,
one might like to simulate data with particular attributes.
For instance, in the MNIST dataset of handwritten digits,
a simple GAN can be utilized to generate fakes, but there
is no control over which handwritten number is simulated.
A cGAN solves the problem by passing a condition, such
as the handwritten digit, through a (potentially trivial)
parameterized function at input, whose output is concatenated with
the usual noise $z\sim P(z)$
and fed into another parameterized function. The combined function
is the cGAN generator $G_\theta$, and the discriminator
proceeds as usual.

In the cases that we study, the condition will be the $h^{11}$
of a K\" ahler metric that we wish to simulate. We will see that this
allows for extrapolation to values of $h^{11}$ that are outside of the
training set. We will be
more precise about the encoding of the condition and the cGAN
architecture when utilizing them in Section \ref{sec:interp_extrap}.

\section{Random Tensor Approximations of String Effective Lagrangians}

\label{sec:random_tensor_approx}

In this section we propose learning random tensor approximations (RTAs) of
low energy Lagrangians that arise from string compactification and exemplify the
idea for K\" ahler metrics on K\"ahler moduli space.

\bigskip

Let us first discuss why learning a RTA is relevant for approximate statistical predictions in
string theory.

 Computing observables $\mathcal{O}(i)$ associated
with a string vacuum $i\in S_\text{vac}$ is facilitated by computing the
Lagrangian $\mathcal{L}_i$ for low energy fluctuations around $i$. For instance, the 
$4d$ renormalizable Lagrangian for self interactions of canonically normalized scalar fluctuations $\phi^a$ around $i$ takes
the form 
\begin{align}
\mathcal{L}_{s,i} = &- \frac12 \partial_\mu \phi^a \partial^\mu \phi^b - M_{ab} \phi^a \phi^b \nonumber \\ &- g_{abc}  \phi^a \phi^b \phi^c - \lambda_{abcd} \phi^a \phi^b \phi^c \phi^d,
\end{align}
where the value of the coupling tensors $M$, $g$, and $\lambda$ are vacuum-dependent.
In general, $\mathcal{L}_i$ also contains fields of other spins and associated coupling
tensors. The couplings are critical in determining $\mathcal{O}(i)$, and therefore an
essential element in making statistical predictions across $S_\text{vac}$ is having
detailed knowledge, and ideally exact computations, for ensembles of coupling tensors.

However, the size of the landscape of vacua and its computational complexity together make
constructing large ensembles of coupling tensors a laborious process. As a concrete
example of the limitations, axion reheating was studied in a large ensemble of string
compactifications with $N$ axion-like-particles (ALPs) in \cite{Halverson:2019kna} and demonstrated to be asymmetric for
all studied values of $N$. Computational limitations required restricting the exact calculations
to $N\leq 200$, despite the fact that $N\sim O(2000)$ is generic in the known ensemble. One 
ulterior motive that we have for proposing the techniques in this work is to be able to estimate
expectations for ALP-cosmology in the large $N$ regime.

Since directly computing large ensembles of coupling tensors is often intractable, it is 
natural to try to simulate them. This is what we mean by a random tensor approximation.
In the language of machine learning, if $x\sim P_d(x)$ is a coupling tensor computed from
some ensemble in string theory, one would like to learn a generative model $G_\theta$ such that 
noise samples $z\sim P(z)$ produce samples $G_\theta(z)$ of an implicit distribution
$P_\theta$ that, after training, yields
\begin{equation}
P_\theta \simeq P_d.
\end{equation}
For instance, if $P_d(x)$ is a distribution on the cubic couplings, one might draw noise
$z$ from a multivariate Gaussian and train $G_\theta$ so that a batches of samples
$z_i \sim P(z)$ yield simulated samples 
\begin{equation}
\hat g_{i,abc} := G_\theta(z_i) \sim P_\theta,
\end{equation}
such that $\hat g$ tensors are indistinguishable from 
$g$ tensors, according to some similarity measure.

We emphasize a crucial difference relative to previous
applications of random matrix theory to the landscape: 
instead of hoping that a well-studied
matrix ensemble approximates string data, we directly
learn the random matrix ansatz using the data. This point
will be discussed further in Section \ref{sec:discussion}.

\bigskip

As a proof-of-principle, we wish to learn a RTA of an ensemble of couplings tensors 
arising in string theory.

Due to their intrinsic interest, we focus on K\" ahler metrics on the K\" ahler moduli space of Kreuzer-Skarke Calabi-Yau
threefolds \cite{Kreuzer:2000xy}. In this case, the tensors are matrices. As a first step, in this paper we will evaluate the K\"ahler metrics
at the apex of the so-called stretched K\"ahler cone, as we wish to focus on learning
the matrix ensemble across a diversity of topologies and extrapolating out of sample. 
In the future it would be interesting to attempt to learn the moduli dependence
of the metric, which would amount to learning random matrix approximations to matrices of polynomial functions.

Let us first discuss the physics of K\" ahler metrics on K\" ahler moduli space. For 
concreteness we will choose to work in type IIB / F-theory. Consider a compactification
of type IIB string theory on a Calabi-Yau threefold $X$. The K\" ahler moduli $T_i$ are the four-dimensional
fields obtain by Kaluza-Klein reduction on $X$ as 
\begin{equation}
T_i =  \int_{D_i} \left(\frac12 J \wedge J + C_4 \right) =: \tau_i + i \theta_i,
\end{equation}
where $D_i$ are $h^{11}(X)$ divisors (four-cycles) that provide a basis for $H_4(X,\mathbb{Z})$
and $J = t_i \omega_i$ is the K\" ahler form, expressed in a basis $w_i \in H^{11}(X)$, and $C_4$ is the Ramond-Ramond four-form.
The kinetic terms for the axion-like particles (ALPs) $\theta_i$ take the form
\begin{equation}
\mathcal{L}_{\theta,\text{kin}} = -M_p^2 K_{ij} \, \partial^\mu \theta^i \partial_\mu \theta^j,
\end{equation}
and similarly for the saxions $\tau_i$. $K_{ij}$ is the metric on K\" ahler moduli space
derived from the classical K\" ahler potential $\mathcal{K} = -2 \,\text{log}\,\mathcal{V}$,
with
\begin{equation}
\mathcal{V} = \int_X J \wedge J \wedge J = \frac16 \, \kappa^{ijk}\, t_i t_j t_k
\end{equation}
the overall volume of $X$ and $\kappa^{ijk}$ the triple intersection numbers on $X$.
The K\" ahler metric is then $K_{ij} = \partial_i \partial_{j} \mathcal{K}$. The
tree-level result receives quantum corrections due to worldsheet instantons,
but the latter are negligible inside the stretched K\" ahler cone, which we will discuss
momentarily.

From the structure of these equations, it is clear that the tree-level $K_{ij}$ is a 
matrix of polynomials in $t_i$. The polynomials themselves have detailed structure
and properties derived from the topology of $X$ and its K\" ahler moduli space. As mentioned,
we will content ourselves to evaluate $K_{ij}$ at points in the moduli space, that is
for specific values of $t_i$. This does introduce a potential source of sample bias into our
studies (though for some applications it is not as severe as one might expect \cite{Halverson:2019kna}). Our goal is to instead focus on diversity across different
Calabi-Yau topologies rather than in the K\" ahler moduli space of a fixed-topology Calabi-Yau.

There is a simple way to see the physical importance of $K_{ij}$. Upon canonical normalization
of the moduli, a change of basis in the fields makes the metric $\propto \delta_{ij}$, eigenvalues of $K_{ij}$ appear in all of the couplings in involving the new fields. The particle physics and cosmology
implications of the ALPs therefore depend critically on $K_{ij}$. For instance, they can
play a crucial role in asymmetric axion reheating \cite{Halverson:2019kna} or couplings to the photon \cite{Halverson:2019cmy}.

Due to the
technical fact that evaluating the inverse K\"ahler metric $K^{ij}$ is computationally
easier, we will actually work with $K^{ij}$, instead of $K_{ij}$. In addition, in order to compare
two K\"ahler metrics corresponding to two different geometries, we will normalize the metric such
that the overall volume of $X$ is one. Extrapolating to other volumes is trivial, as the metric
is a homogeneous function of the moduli.

With the above motivation and context in mind, for the remainder of the paper we will
focus on learning random matrix approximations to $K^{ij}$.

\bigskip

To do so, we must have an ensemble to learn from.
The algorithm we use to generate data is as follows:
\begin{algorithm}[H]
    \caption{Generate ensemble of K\"ahler metrics.}
\begin{algorithmic}[1]
    \Require Fixed value of $h^{11}$, set $S_\text{poly}$ of reflexive $4d$ polytopes
     with that value of $h^{11}$.
     \For{polytope $P \in S_\text{poly}$}
        \State FRST $\gets$ pushing triangulation of $P$.
        \State A $\gets$ ToricVariety(FRST).
        \State X $\gets$ generic anticanonical hypersurface in A.
        \If{$h^{11}(A)=h^{11}(X)$}
            \State $J \gets$ parameterized K\"ahler form.
            \State $\mathcal{V} \gets \frac16 \int_X J \wedge J \wedge J$.
            \State $\mathcal{K} \gets -2\text{log}(\mathcal{V})$.
            \State $D_i \gets$ toric divisor, where $i=1,\dots, h^{11}+4$.
            \State $C_{ij} \gets D_i \cdot D_j\cdot X, \,\,\, \forall i,j$.
            \State $a \gets$ point in  K\" ahler moduli space such that vol$(\sum_{i,j} C_{ij})$ is minimized 
            subject to the stretched K\"ahler cone condition vol$(C_{ij})>1\,\, \forall i,j$.
            \State $\tilde{a} \gets$ rescale a such that $\mathcal{V} = 1$.
            \State  $K^{ij} \gets (\partial_i \partial_{j} \mathcal{K})|_{\tilde{a}}^{-1}$.
            \State save $K^{ij}$ and its eigenvalues, which are $>0$.
        \EndIf
     \EndFor
\end{algorithmic}
\end{algorithm}
\noindent The region in K\" ahler moduli space satisfying $\text{vol}(C_{ij})\geq 1$ for all $i,j$ is the so-called stretched K\" ahler cone; the point $a$ is known as its apex. Both were
introduced in \cite{Demirtas:2018akl}.

\begin{table}[t]
\begin{tabular}{c|ccccc|}
 $h^{11}$ & $10$ & $20$ & $30$ & $40$ & $50$ \\ \hline
\# Favorable & $9282$ & $5793$ & $4222$ & $5517$ & $4899$ \\ \hline
\end{tabular} \\ \bigskip
\begin{tabular}{c|cccccc|}
 $h^{11}$ & $20$ & $22$ & $24$ & $26$ & $28$ & $30$  \\ \hline
\# Favorable & $5793$ & $4936$ & $5152$ & $3981$ & $4074$ & $1722$ \\ \hline
\end{tabular}
\label{tab:favorable}
\caption{For each $h^{11}$, the number
of favorable Calabi-Yau hypersurfaces associated to $10,000$ toric ambient spaces obtained
by pushing triangulations of $4d$ reflexive polytopes. Top and bottom are values of $h^{11}$
utilized in fixed $h^{11}$ and interpolation / extrapolation experiments, respectively.}
\end{table}

For each $h^{11}\in \{10,20,22,24,26,28,30,40,50\}$ we studied the first $10,000$ polytopes
from \cite{KSdata}. Utilizing standard toric geometry packages in Sage, taking the fine regular star triangulation (FRST) gives
an ambient space $A$, and the associated Calabi-Yau hypersurface $X$ is called favorable if
$h^{11}(A) = h^{11}(X)$. Since we only study favorable cases, we simple refer to $h^{11}:= h^{11}(X) = h^{11}(A)$. The number of favorable geometries is given in Table \ref{tab:favorable}, split according to values of $h^{11}$ utilized in two different
types of experiments.
Further details for data generation can be found in the GitHub repository \cite{git_repo}.

We will use this data in two different types of experiments, designed to test performance
at fixed $h^{11}$, as well as the ability to interpolate or extrapolate out of sample,
i.e. to values of $h^{11}$ not utilized in training.

\bigskip
We will also overcome a common problem with generative models
based using the nature of K\" ahler metric data. The problem is
that it is often unclear how to evaluate the performance of
$G_\theta$. For instance, if $G_\theta$ is trained to 
provide deep fakes of human faces, the performance could
be evaluated by asking humans to determine whether a 
set of samples if real or fake. However, this rather brittle process is expensive and slow.
One would like a numerical figure-of-merit that may
be easily computed and utilized to compare real data against fake data.

In our case, we will utilize the fact that our ``images" are matrices that
appear in low energy effective Lagrangians in
string theory, the eigenvalue
spectrum of which carries physical information. By contrast,
it doesn't make sense to study the eigenvalue spectrum
of human faces. Our figure-of-merit
will be the distribution of $\text{log}_{10}$ of the eigenvalues
of $K^{ij}$, and specifically we will study how the Wasserstein
distance between the real and fake log eigenspectra changes as
the GANs are trained. That is, at fixed
$h^{11}$ we produce $\Ng$ simulated inverse K\" ahler metrics
$K^{ij}$, compute the log eigenspectrum of those samples, and 
compute the Wasserstein distance relative to the 
the test ensemble of real $K^{ij}$. (The test ensemble of
K\" ahler metrics is the complement of the 
training ensemble inside the set of K\" ahler metrics 
on the favorable geometries in Table \ref{tab:favorable}).  One
expects the eigenspectrum distance to decrease during training.

There is an obvious potential confusion that we would
like clarify: this Wasserstein distance of log eigenspectra
that is our figure-of-merit
is completely separate from the Wasserstein distance
implicit in WGANs. The latter is an approximate
Wasserstein distance between
$P_d$ and $P_\theta$, which is intractable due
to the high dimension of the distributions and
is therefore estimated using Kantorovich-Rubinstein
duality. The former is simply the Wasserstein
distance of the one-dimensional log eigenvalue distributions,
and is readily computed using SciPy. \emph{Specifically,
we do not train on the log eigenvalue distribution.}

\subsection{K\" ahler Metric Simulation at Fixed $h^{11}$}
\label{sec:fix_h11}

We first learn random matrix approximations of K\" ahler metrics
at fixed values of $h^{11}$.  The parameters available to our experiments are:
\begin{center}
\begin{tabular}{c|c|}
Param. & Description \\ \hline
Model & GAN, WGAN, DCGAN, or DCWGAN  \\
$h^{11}$ & Hodge number of $h^{11}(X)$ \\
$N_\text{geom}$ & \# of geometries $X$ used in training\\
$n_z$ & \# of draws from $\mathcal{N}(0,1)$ at input \\
$N_\text{batch}$ &  batch size \\
$N_\text{crit}$ & \# of WGAN critic loops (if applicable) \\
$\alpha$ & learning rate for RMSProp \\ \hline
\end{tabular}
\end{center}
\smallskip
where the GAN and WGAN in the model type denote the GAN loss and Wasserstein 
GAN loss introduced in Section \ref{sec:generative_intro}. Both algorithms require
a generator network and a discriminator network,
\begin{align}
G_\theta&: \bR^{n_z} \to \bR^{h^{11}} \times \bR^{h^{11}} \nonumber \\
D_w&: \bR^{h^{11}} \times \bR^{h^{11}} \to D_T,
\end{align}
where in the Wasserstein GAN case the discriminator is often called the critic.
$D_T$ is the discriminator target; for the GAN it is $[0,1]$, and for the WGAN it is
$\mathbb{R}$.
 The presence of DC in the model type
denotes a deep convolutional architecture; otherwise it is a fully connected
feed-forward network. Further details of the architecture
can be found in the repository \cite{git_repo}.

In the case of a Wasserstein GAN, $N_\text{crit}$ is the number
of batches the critic is trained on for each generator training batch. This
parameter is crucial because, as discussed, the Wasserstein GAN requires a strong
critic. If performance is poor, it may be due to a weak critic, which can be solved
by increasing $\Nc$.

\medskip
We run the first batch of experiments with fixed 
\begin{equation}
(\Ng,\Nb,\Nc)=(2500,64,5),
\end{equation}
models varying across the listed types, and
\begin{align}
h^{11}&\in\{10,20,30,40,50\}, \nonumber \\ 
\alpha&\in\{5\times 10^{-5}, 5 \times 10^{-6}\}, \nonumber \\
n_z&\in \{5,15,25,50\},
\end{align} for a total of $160$ experiments, $1000$ epochs each. 

\begin{figure}[t]
\includegraphics[width=.98\columnwidth]{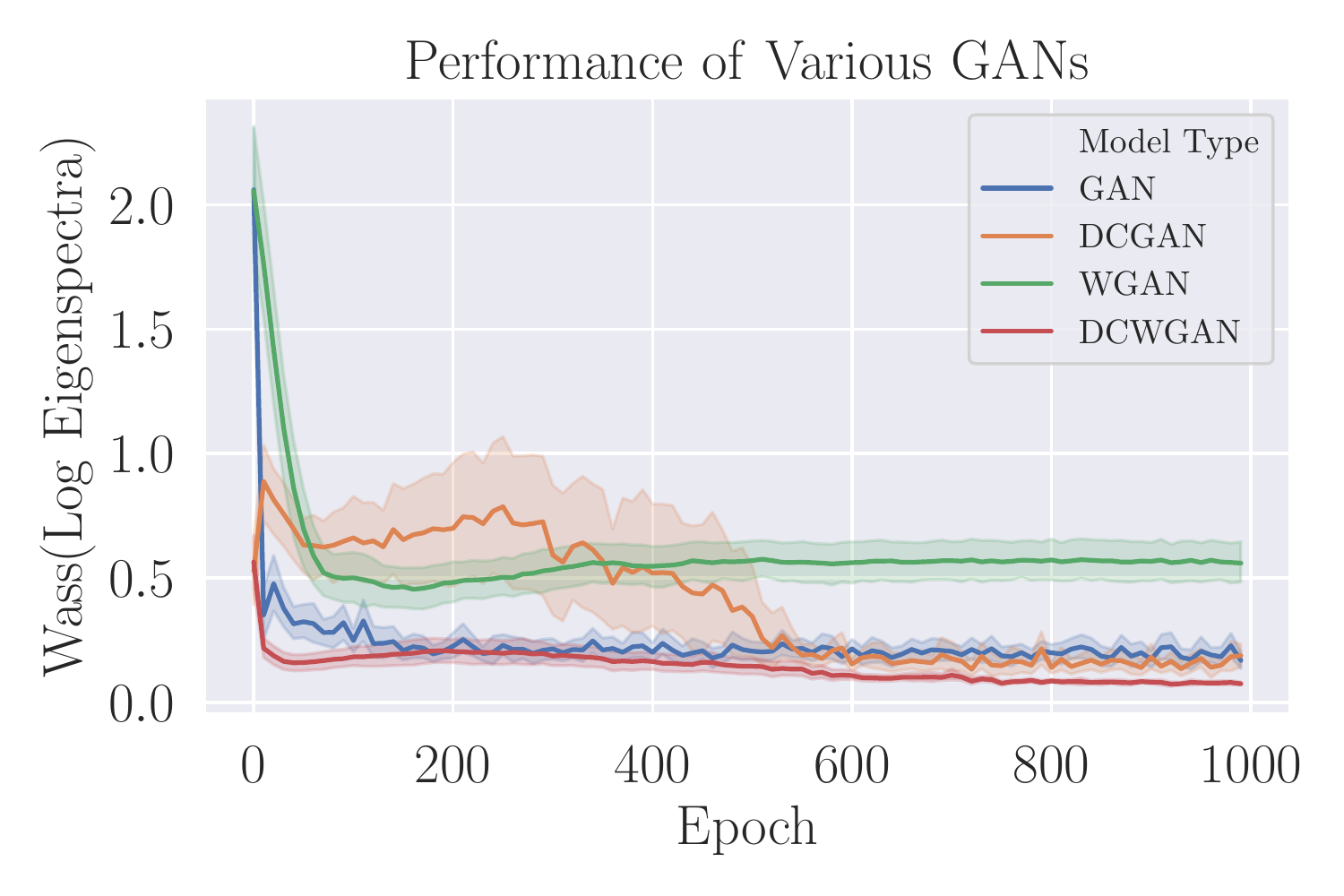}
\hspace{.5cm}
\includegraphics[width=.98\columnwidth]{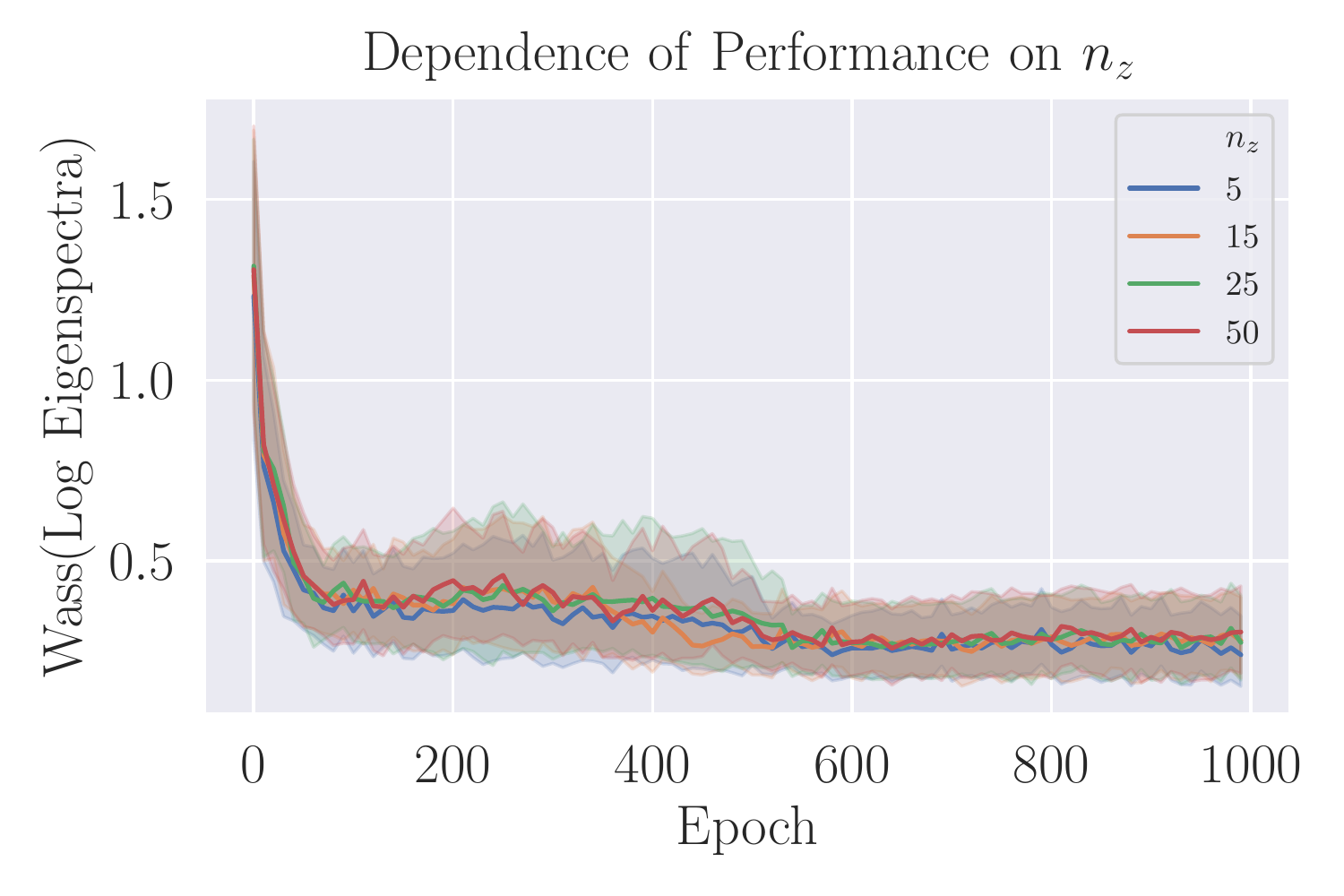}
\caption{Performance of fixed $h^{11}$ experiments with $\alpha = 5 \times 10^{-6}$, dependent upon model type (left) and 
the number of Gaussian draws $n_z$ (right), with $95\%$ confidence intervals. \emph{Top:} a Wasserstein GAN with deep convolutional architecture
gives the best performance and fastest training. \emph{Bottom:} high accuracy simulation is achieved with little variance
across the number of Gaussian draws, even with $n_z = 5 \ll h^{11} \in \{10,20,30,40,50\}$.}
\label{fig:fixed_h11}
\end{figure}

Results are presented in
Figure \ref{fig:fixed_h11}, where we have focused
on the $\alpha = 5\times 10^{-6}$ since the lower learning
rate decreases noise and clarifies the result. On top, we see that performance, as measured
by the Wasserstein distance between the real and fake
log 
eigenspectra, depends critically on the model type.
A DCWGAN clearly performs best. 
This is not a surprise, as the Wasserstein GANs 
and / or deep convolutional architecture often improve
GAN training. On the  bottom, we see the the
performance effectively does not depend on $n_z$ in the
ranges we have chosen; note that performance does go
down for $n_z=1$, however. This point is worthy of
significant discussion, see Section \ref{sec:discussion}.

\medskip
We also ran another experiment to aid in visualizing
the results with respect to the actual images and the
converging eigenspectra. 
The experiment is a DCWGAN with 
\begin{align}
h^{11} &= 10,\,\,\, \Ng = 2500, \,\,\, n_z = 5 \nonumber \\
\Nb &= 64,\,\,\, \Nc = 5,\,\,\, \alpha = 2.5\times 10^{-6}.
\end{align}
The progression of log eigenspectra and image representations
during training are presented in 
Figures \ref{fig:spectrum_change}
and \ref{fig:image_change}, respectively. In the former,
the eigenspectra are seen to converge to good agreement.
The plots also serve as a heuristic gauge for 
what Wasserstein distances of log eigenspectra correspond
to good agreement between simulation and real data. To the naked eye,
distances of $\lesssim .2$ have good agreement,
whereas the distance $.94$ at epoch $0$ demonstrates
a poor model.
In Figure \ref{fig:image_change}, samples that were blurry and faint
at early times increase in sharpness and contrast during
training,
looking increasingly realistic to the naked eye.

\subsection{Interpolation and Extrapolation in $h^{11}$ with
Conditional GANs}

\label{sec:interp_extrap}

Since we would like to be able to reliably simulate string
data in regimes where exact computation is intractable, we
now study whether GANs for string data are able to
interpolate and extrapolate. 
Specifically, we study whether it is possible to interpolate or extrapolate in $h^{11}$,
relative to the $h^{11}$ values of the training samples.

A priori this seems like a bad idea, 
because extrapolating out of sample is in general intractable, but in special cases it may be possible if the data is 
highly structured. This is often the case in string theory, and
in the data that we study the structural relationship is
due to topological transitions that change $h^{11}$. 
We will speculate about this further in Section \ref{sec:discussion}.

\bigskip
Since we wish to interpolate and extrapolate,
the techniques must differ in crucial ways
from those of Section \ref{sec:fix_h11}, though
many of the parameters are the same. 

First, we must introduce conditions, so that the input to the generative model is not only noise
$z\sim P(z)$, but also some information about the nature of the sample
we wish to generate. For us, it is $h^{11}$ that we wish
to pass as a condition. We one-hot encode\footnote{A one-hot encoding of an integer
$i$ represents $i$ by the unit vector $e_i \in \mathbb{Z}^k$, where there are $k$ different
allowed values of $i$.} the value of $h^{11}$ and 
pass it through a function:
\begin{equation}
C_{\phi}: \mathbb{Z}^k \to \mathbb{R}^l
\end{equation}
where $l$ is a hyperparameter and $C$ may have non-linearities.
The noise input $z\sim P(z)$ is concatenated with $C_{\phi}(c)$ for $c\in\mathbb{Z}^k$
and passed as input to
\begin{equation}
N_\varphi:\bR^{l+n_z} \to \bR^{\text{max}\, h^{11}} \times \bR^{\text{max}\, h^{11}},  
\end{equation}
which together form the generator
\begin{equation}
G_\theta: \mathbb{Z}^k \times \mathbb{R}^{n_z} \to \bR^{\text{max}\, h^{11}} \times \bR^{\text{max}\, h^{11}}
\end{equation}
via $G_\theta(c,z)=N_\varphi(C_\phi(c),z)$, so that the parameters $\theta$ are the union
of $\varphi$ and $\phi$.
For us, $C_\phi$ is a fully-connected layer with LeakyReLU activation and $N_\varphi$ is
effectively one of the $G_\theta$ of Section \ref{sec:fix_h11}, together with some additional zero-padding
since the data is not uniform, due to varying $h^{11}$. For architecture details, see the
repository \cite{git_repo}.

Second, we must state the relationship between interpolation,
extrapolation, and the conditions. 
If the values of $h^{11}$ utilized during training
and testing are
\begin{align}
h^{11}_\text{train} &= \{20,22,28,30\} \nonumber \\
h^{11}_\text{test} &= \{20,22,24,26,28,30\},
\label{eq:interp}
\end{align}
then the set 
$h^{11}_\text{test}\setminus h^{11}_\text{train}=\{24,26\}$
means that we test also for $h^{11}$ values that are in
between the training values; this is interpolation.
Similarly, if 
\begin{align}
h^{11}_\text{train} &= \{20,22,24,26\} \nonumber \\
h^{11}_\text{test} &= \{20,22,24,26,28,30\},
\label{eq:extrap}
\end{align}
then accurate predictions for $h^{11}_\text{test}\setminus h^{11}_\text{train}=\{28,30\}$
corresponds to extrapolation.  Again our figure-of-merit
is the Wasserstein distance of the log eigenspectra of
$K_{ij}$, but now there are six comparisons, one for
each $h^{11} \in h^{11}_\text{test}$, two of which
do not appear in the train set. We are testing
if the cGAN simulate K\" ahler metrics  for values of $h^{11}$ not involved in training.

\begin{figure*}[t]
\includegraphics[width=.98\columnwidth]{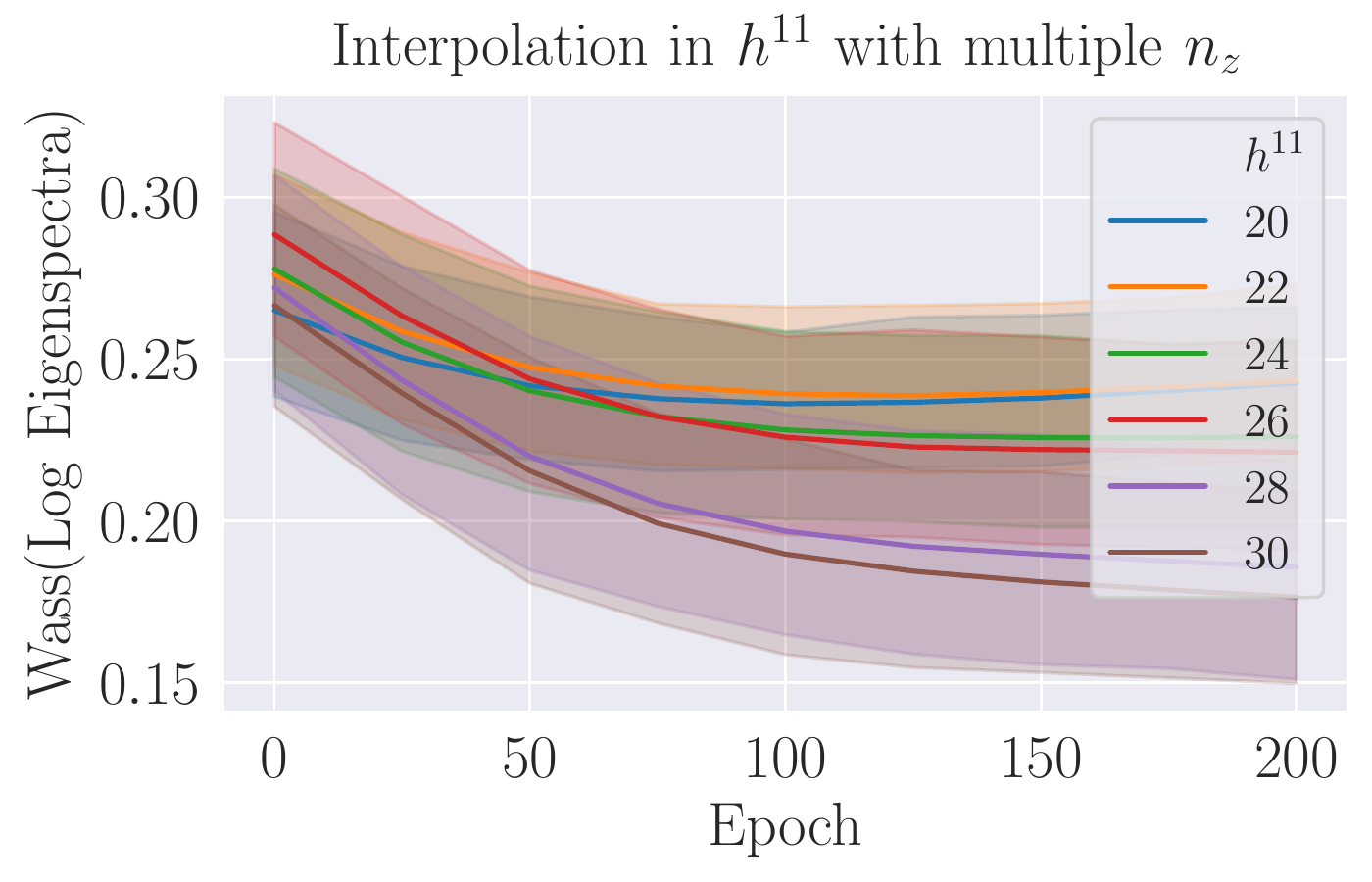}
\hspace{.5cm}
\includegraphics[width=.98\columnwidth]{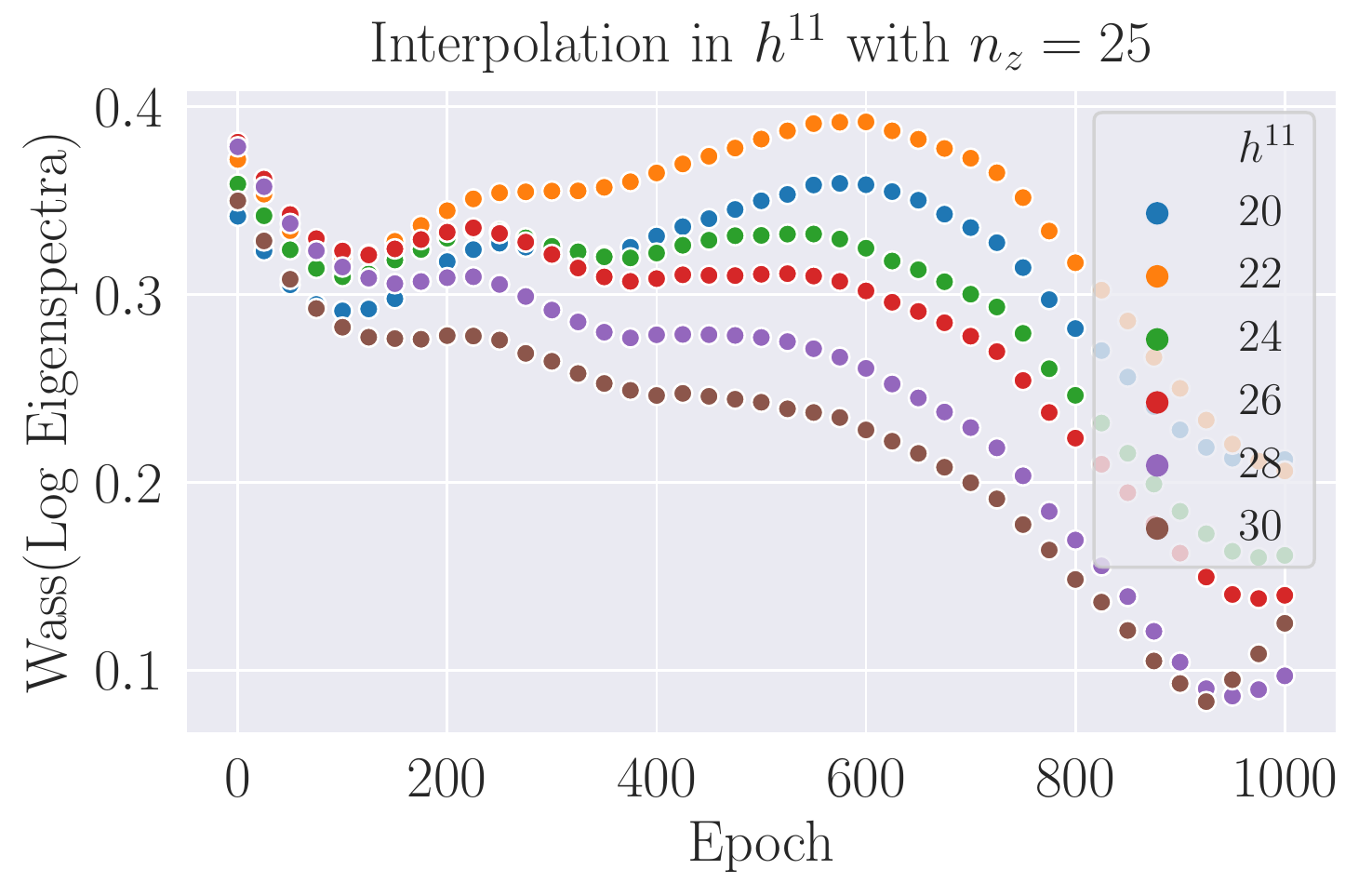}  \\ 
\includegraphics[width=.98\columnwidth]{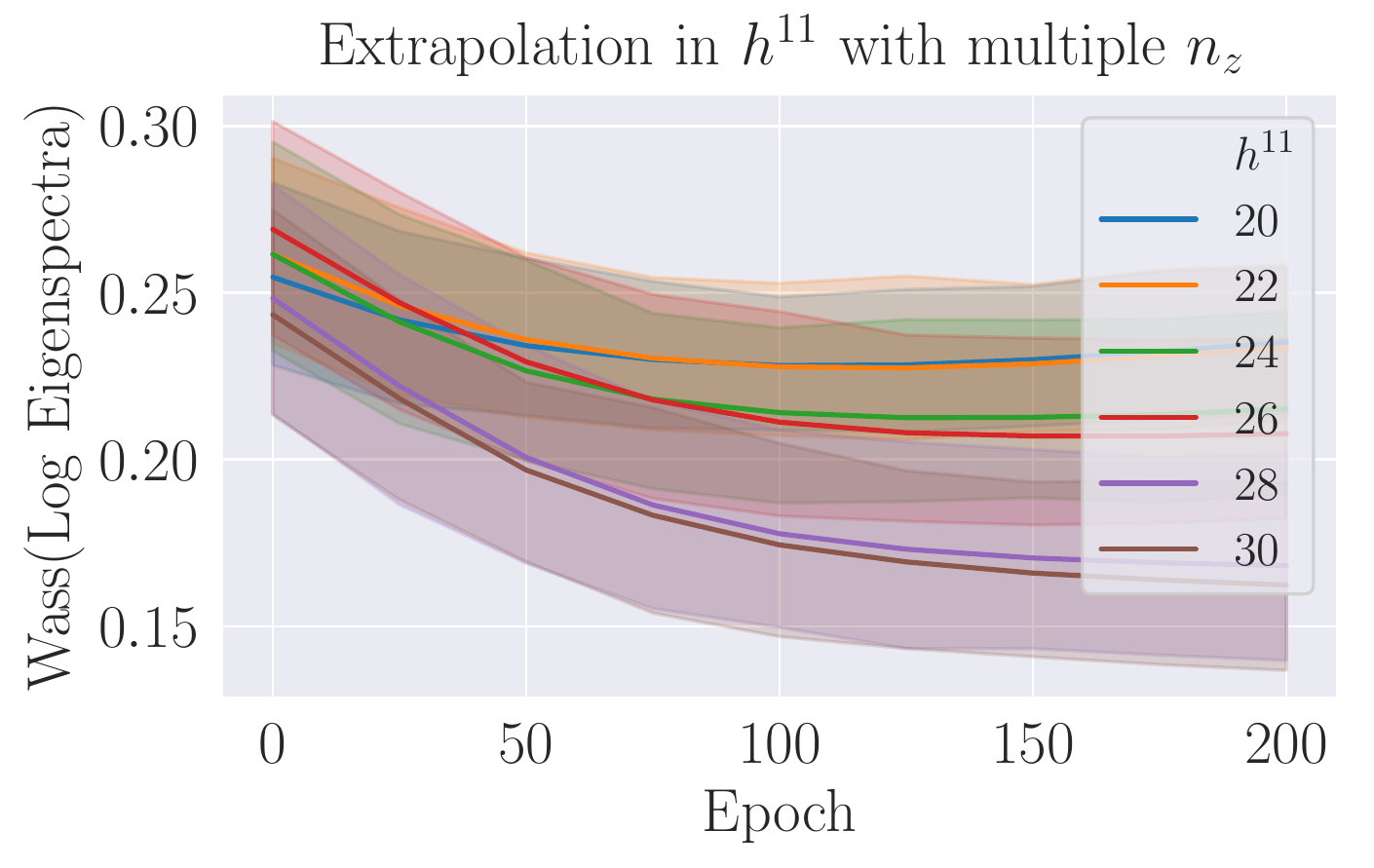}
\hspace{.5cm}
\includegraphics[width=.98\columnwidth]{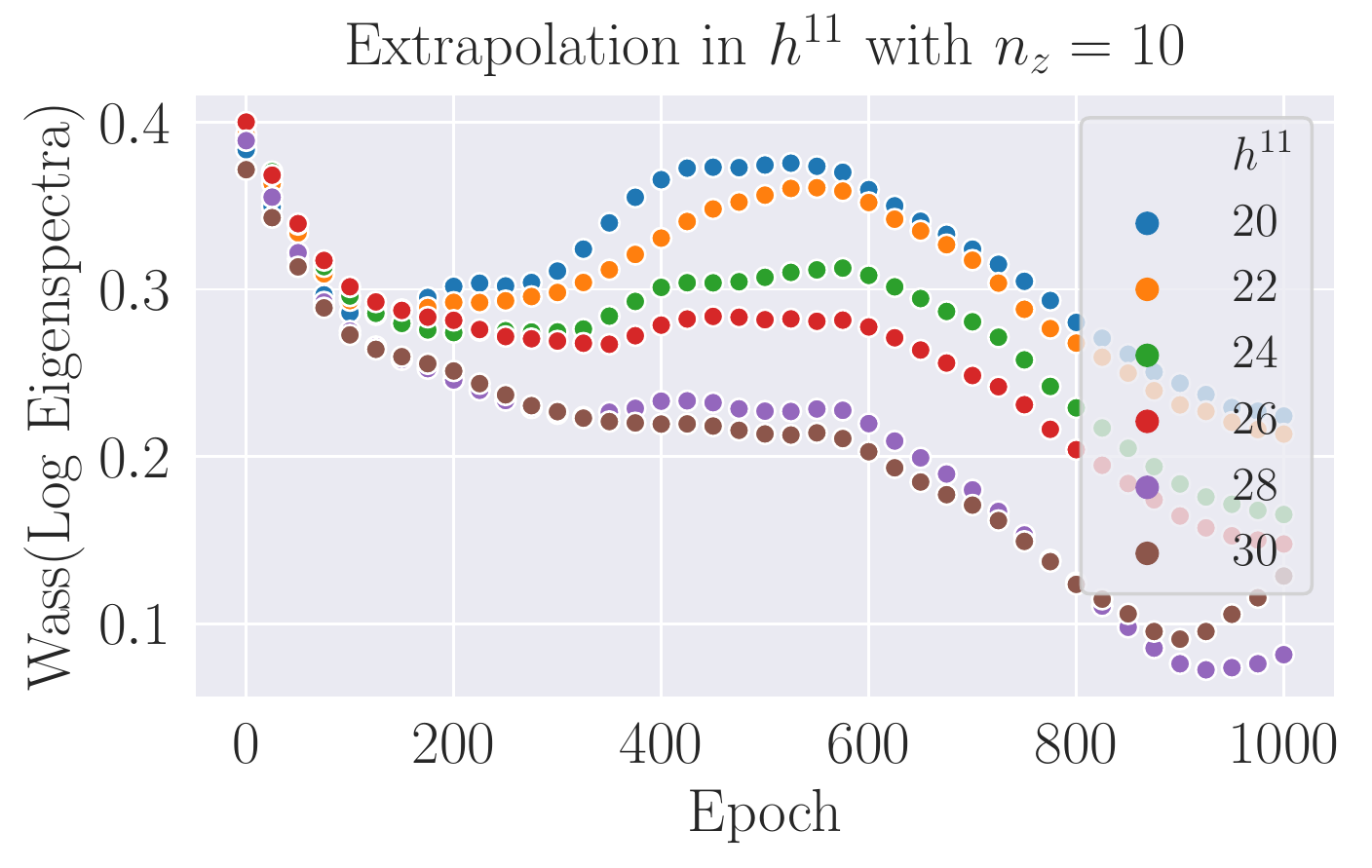} 
\caption{Performance of interpolation (top) and and extrapolation (bottom) experiments,
as a function of the parameter $h^{11}$ supplied as a condition to the GAN. \emph{Left:} 
Mean performance and 95\% confidence intervals. \emph{Right:} Illustrative single experiments
with stated $n_z$ and learning rates of $10^{-7}$. }
\label{fig:interp_extrap}
\end{figure*}

\bigskip

Given the success of the Wasserstein DCGAN in simulating K\" ahler metrics
at fixed $h^{11}$, we promote this model alone to become a conditional GAN, so that the full model we study
for interpolation and extrapolation is 
a conditional deep convolutional Wasserstein GAN. This means that $N_\varphi$ is a 
deep convolutional network and the associated generator $G_\theta$ and $D_w$ are trained
as a Wasserstein GAN. The parameters are
\begin{center}
\begin{tabular}{c|c|}
Param. & Description \\ \hline
$h^{11}_\text{train}$ & $h^{11}$ values of training set \\
$h^{11}_\text{test}$ & $h^{11}$ values of test set \\
$N_\text{geom}$ & $\#$ geometries $X$ used in training per $h^{11}$\\
$n_z$ & \# of draws from $\mathcal{N}(0,1)$ at input \\
$N_\text{batch}$ &  batch size \\
$N_\text{crit}$ & \# of WGAN critic loops (if applicable) \\
$\alpha$ & learning rate for RMSProp \\
$l$ & width of latent layer encoding for $h^{11}$\\ \hline
\end{tabular}
\end{center}
In our experiments, $h^{11}_\text{train}$ and $h^{11}_\text{test}$ as chosen as in \eqref{eq:interp}
and \eqref{eq:extrap} for interpolation and extrapolation, 
and  we take $k=\text{max}\, h^{11}=\text{max}(h_\text{test}^{11})$. Furthermore we take
\begin{equation}
(\Ng,\Nb,\Nc,\alpha)=(2500,64,5,10^{-7}),
\end{equation}
and
\begin{align}
n_z&\in \{10,25,50,100\},
\end{align} for a total of $4$ different experiment types for interpolation, and $4$
for extrapolation. We found that these experiments, perhaps due to the complexity
of the input and architecture, lead to more noise, and we therefore ran each of these
experiments $10$ times to build statistics.

Results are presented in Figure \ref{fig:interp_extrap}. From the mean performance
plots, we see clear evidence of learning across all values of $h^{11}\in h^{11}_\text{test}$,
with the trend that learning is a bit modest for small $h^{11}$. The decreasing performance with decreasing $h^{11}$ is likely due to the fact
that the smaller the value of $h^{11}$, the more zero-padding is necessary. For instance,
a metric with $h^{11}=20$ has $400$ entries, but since $30\in h^{11}_\text{test}$ it is zero-padded
such that it has $500$ more zeroes than every metric with $h^{11}=30$. This effect is almost
certainly solvable with a more clever architecture that allows for non-uniform data. Nevertheless, learning
occurs for all values of $h^{11}$.

Two specific experiments are also presented, to demonstrate trends that are common
in many of the experiments. Specifically, experiments that start with a large
Wasserstein loss often have significant learning in the first $O(200)$ epochs, but then
experience a bump that decreases the performance, particularly at smaller $h^{11}$. This is sometimes overcome with additional learning at late times that leads to the best results,
as demonstrated on the RHS of Figure
\ref{fig:interp_extrap}. In some cases the experiments start with relatively low Wasserstein loss,
in which case significant learning does not necessarily occur.

\bigskip
Most notably, as is the point of this section, we emphasize that these generative models demonstrate the ability to interpolate
and extrapolate to metrics at values of $h^{11}$ that were not involved in training.
Specifically, in the top two plots of Figure \ref{fig:interp_extrap} the learning
associated with the $h^{11}\in \{24,26\}$
data demonstrates the ability of the cGAN to interpolate, while the bottom two plots
exhibit extrapolation due to the learning associated with the
$h^{11}\in \{28,30\}$ data. Reasons that we did not push the technique further will
be addressed in Section \ref{sec:discussion}, but we consider this a successful proof-of-principle of the ability
of generative models to exhibit some extrapolation on string theory data, perhaps
due to structural topological relationships between geometries.

Our GAN approach allows for fast simulation. The trained conditional DCWGAN provides a speedup of generation of K\" ahler metrics at $h^{11} = 30$ by a factor of about 250, compared to the current leading pipeline\footnote{We thank Mehmet Demirtas for performing a computation of K\" ahler metrics to which we can compare our results.} from polytope to effective Lagrangian~\cite{mehmet}, and so yields a large speedup at fixed $h^{11}$. Importantly, while the pipeline in~\cite{mehmet} will have at least a polynomial-time slowdown with $h^{11}$, the speed of the conditional DCWGAN is fixed across all $h^{11}$, since it is input to a fixed trained neural network. Therefore, if one can actually use this technique to extrapolate to large $h^{11}$, the conditional DCWGAN will likely provide a means to sample effective Lagrangians at large $h^{11}$ where no other technique will be fast enough to provide useful statistics.

\section{Discussion}
\label{sec:discussion}

In this paper we have introduced a new approach to making statistical
predictions in string theory.
We proposed the use of deep generative models,
a class of techniques in machine learning that train a generator function (deep neural network)
$G_\theta$ to convert draws from a distribution $P(z)$ to draws from
a distribution $P_\theta$ that approximates a data distribution $P_d$.
Specifically,
we utilized generative adversarial networks (GANs), but this is simply
an instantiation of the broader idea, and it is worth exploring other
possibilities as well.

To see the utility of such techniques in string theory, consider what one would  do
in the presence of an all-powerful oracle with perfect knowledge of the string landscape.
The oracle knows the full set of vacua $S_\text{vac}$ and the cosmological probability
distribution $P(i)$ on it. It can efficiently sample $P(i)$ and compute any observable $O(i)$ for any $i\in S_\text{vac}$.
Then there is no obstacle to making statistical predictions: one simply uses the oracle to
collect enough samples from $P(i)$, computes ensemble averages of observables, and compares to experiment.

Of course, this oracle is rather futuristic. We currently only know subsets of $S_\text{vac}$, albeit very large ones,
and despite some progress there is still much to be understood about dynamical and anthropic contributions to $P(i)$.
Furthermore, in some classes of vacua it is not known how to compute some basic observables, or it is simply inefficient,
sometimes due to running up against instances of NP-hard problems. Even a weaker oracle that only knows $S_\text{vac}$,
$P(i)$, and how to compute observables has a serious problem: sampling is non-trivial, yet crucial to making
statistical predictions.

\bigskip

By learning a distribution $P_\theta$ that approximates a data distribution $P_d$ and generating samples from
$P_\theta$, generative models offer the possibility of trading some error for efficient sampling.
If the error is sufficiently small and/or controllable, this provides a useful means for making
\emph{approximate} statistical predictions in string theory. That is the central conceptual idea in this paper.

There is a down-to-earth application of this idea that we explored. For vacua whose low energy
fluctuations admit a Lagrangian description, learning to approximately sample them corresponds to learning
a random tensor approximation (RTA) for the couplings in the Lagrangian.
For two-index couplings, this is simply learning
a random matrix approximation (RMA).

This is markedly different from previous applications  of random matrix
theory (RMT) in or inspired by the string landscape \cite{Marsh:2011aa, Chen:2011ac,*Pedro:2013nda,Long:2014fba,*Achucarro:2015kja,*Pedro:2016jyd,*Pedro:2016sli, Bachlechner:2012at, Bachlechner:2014rqa}. There, it was often the case that well-studied random matrix ensembles were studied at large $N$ (number
of fields, cycles, etc) and universality yielded physical implications. However, it is not clear a priori why
such ensembles should have anything to do with string theory, which exhibits structures that may violate assumptions of
certain RMT ensembles. In fact, observables in known ensembles of the type IIB theory compactified on Calabi-Yau manifolds
deviate \cite{Long:2014fba} from the expectations of canonical RMT ensembles.

\bigskip

Instead, generative models offer a means of \emph{learning} RTAs of string effective Lagrangians.

We exemplified the idea in the case of K\" ahler metrics on the K\" ahler moduli space of Calabi-Yau manifolds. Such metrics were generated for thousands of Kreuzer-Skarke
Calabi-Yau threefolds
at various values of $h^{11}$, which served as training data from which to learn random tensor approximations.
We utilized multiple different types of GANs, which differ according to their loss functions (a normal GAN
versus a WGAN) and architecture (fully connected feedforward versus convolutional). In each case, the GAN generator is
a deep neural network $G_\theta$ that produces simulated K\" ahler metrics as $G_\theta(z)$, where $z$ is
a vector of noise of dimension $n_z$ with entries drawn from the Gaussian distribution $\mathcal{N}(0,1)$.

Unlike in many applications of GANs, we have a natural
figure-of-merit by which to judge the learning process: the Wasserstein distance between the log eigenvalue
distributions of the real and fake K\" ahler metrics. Given $N$ real K\" ahler metrics arising
from Calabi-Yau manifolds and $N$ fake K\" ahler metrics $G_\theta(z)$, with $N$ sufficiently large,
we compute the log eigenspectrum. A bad RMA of K\" ahler metrics will have significant mismatch between the
log eigenspectra. If learning is occurring as $G_\theta$ is trained then they should increasingly overlap,
which we measure with the Wasserstein (a.k.a. earth-mover) distance; as discussed, this use of Wasserstein
distance is fundamentally different from that of the WGAN.

\bigskip
We performed two classes of experiments that demonstrate learning of RMAs of K\" ahler metrics.

\textbf{Fixed $h^{11}$ results:} In Section \ref{sec:fix_h11}, we trained GANs at fixed values of $h^{11}\in \{10, 20, 30, 40, 50\}$. In all cases, the Wasserstein
distance on log eigenspectra decreases significantly during training. We found that a Wasserstein GAN
with deep convolutional architecture (DCWGAN) significantly outperforms the other GAN types that we tried.
When viewing the K\" ahler metrics as grayscale images, we found that at early times some of the images were
faint with low contrast relative to the real K\" ahler metrics. This improved upon training; i.e., some aspects
of learning can be seen with the naked eye.

Perhaps the most important result for the fixed $h^{11}$ experiments is that taking different values of
$n_z\in \{5,15,25,50\}$ did little to affect performance, at least with respect to the Wasserstein distance
on the log eigenspectra. This is rather remarkable: despite the disparate $h^{11}$ values and thousands of geometries
utilized for each, the neural network is able to generate matrices whose eigenspectrum resembles the Calabi-Yau
data using only $5$ Gaussian draws,\footnote{Note that one can try to take this too far: performance goes down significantly
  for $n_z=1$, for instance.} where $5\ll h^{11}$. This suggests that the so-called ``data manifold'' is of relatively
small dimension, demonstrating implicit correlations.

It is worth commenting further on this data manifold in light of the difference between RMAs and the well-studied random
matrix ensembles previously applied in the string literature. For instance, it might be considered natural to use the
Wishart ensemble to model K\" ahler metrics, since its matrices are also positive definite.\footnote{A Wishart matrix is of the form
  $A^T A$, where the matrix $A$ has its entries independently and identically distributed (i.i.d.) according to a Gaussian distribution.}
However, due to the $N^2$ i.i.d. entries the Wishart ensemble has dimension $N^2$ support in the space of $N\times N$ matrices. This is
clearly different from K\"ahler metrics on K\"ahler moduli space (with $N=h^{11}$), which despite being $N \times N$ matrices
nevertheless only depend on $h^{11}$ variables: the K\" ahler moduli. On general grounds, then, one should not expect the
Wishart ensemble to be a good approximation to K\" ahler metrics.

In this light, we revisit the fact that even $n_z=5$ GANs yielded good simulations of K\" ahler metrics. From the
fact that they are functions of K\" ahler moduli space, one expects that $n_z=h^{11}$ draws should suffice, given
a sufficiently expressive neural network, but in fact $n_z\ll h^{11}$ seems to do rather well. Clearly
this cannot be exactly true, since the exact (rather than approximated) ensemble of tree-level K\" ahler metrics
depends explicitly on a manifold of dimension $h^{11}$, the K\" ahler moduli space. This deserves
further thought, and we will return to it after summarizing another result.

\textbf{Extrapolation in $h^{11}$ results:} In the second class of experiments, studied in Section \ref{sec:interp_extrap}, we studied whether the GAN had the
ability to interpolate or extrapolate out of sample. This is of interest because computational complexity often limits
exact computations to moderate $N$ regimes (see, e.g., the ALP example in the text), despite the fact most vacua are
expected to live at large $N$. Clearly it would be beneficial if a GAN could simulate string data at large $N$, if
exact computations are not available there. While a priori one should be skeptical of such extrapolation, it may perhaps
be possible if the data is highly structured, as it often is in string theory.

To attempt interpolation and extrapolation, we used a conditional GAN (cGAN) with Wasserstein loss function and deep
convolutional architecture; a cDCWGAN, putting the pieces together. The key difference in a conditional GAN is that the input
is not simply noise drawn from some distribution, but also a condition that dictates what \emph{type} of sample to generate.
For instance, in generating handwritten digits, one might wish to have the ability to choose whether to generate a seven
or a nine. For us, we passed $h^{11}$ as a condition, so that the GAN learns to simulate K\" ahler metrics at a chosen
value of $h^{11}$. Interpolation (extrapolation) then corresponds to the accurate generation of K\" ahler metrics (as
measured by Wasserstein distance of log eigenspectra) for values of $h^{11}$ in between (larger than) the values of $h^{11}$
of the real K\" ahler metrics used in training. Specifically, in the interpolation experiments we trained at $h^{11} \in \{20, 22, 28, 30\}$
and tested for those values, as well as the interpolated values $h^{11}\in \{24,26\}$. For extrapolation, we trained at
$h^{11}\in \{20,22,24,26\}$ and tested at those values, and also the extrapolated values $h^{11}\in \{28,30\}$.

The result is that the GAN learned to generate K\" ahler metrics at values of $h^{11}$ not utilized in training, i.e. it was
able to both interpolate and extrapolate. There was decreased performance for smaller $h^{11}$, almost certainly
correlated with increased amounts of zero-padding for smaller $h^{11}$; this can likely be overcome by utilizing architectures
that allow for non-uniform data. We also point out that we did not attempt to extrapolate further in $h^{11}$, since
as $h^{11}$ increases the eigenvalue distribution for K\" ahler metrics becomes bimodal, and thus far we have found it
difficult to model the second mode, though we expect this is doable with future advances. It would also be interesting to
understand geometric origin of the second mode, which may be related to qualitative changes (such as increasing numbers of facet interior points)
of the associated reflexive polytopes as $h^{11}$ increases.

\bigskip
\textbf{Concluding comments.} Following our proposal for making approximate statistical predictions in
string theory, we have presented concrete results that demonstrate
the ability of generative models to simulate string data. Though we specifically used GANs to simulate
K\" ahler metrics, there is no clear obstruction preventing the use of other generative models or studying
other types of data, including structures in formal theory that may not be as relevant for the landscape. For instance,
one could utilize normalizing flows, which not only give the ability to generate samples, but also allow
for the computation of the
probability of the sample in $P_\theta$, due to the generative model being an invertible neural network.

The neural networks performed surprisingly well in at least two ways. First, with very few random draws, $n_z=5\ll h^{11}$, they
were able to efficiently simulate K\" ahler metrics at fixed $h^{11}$. Second, the
conditional GAN was able to extrapolate, simulating K\" ahler metrics at values of $h^{11}$ not seen during training. From a machine learning perspective, these facts suggest the presence of structure that is making learning possible.

Perhaps it is the highly structured and relational nature of string data the makes it learnable. Not
only is a single data point typically accompanied by significant structure, such as the topological and geometric
information carried by a fixed string vacuum, but these data points are related to one another by deformations
or discrete operations in a mathematically rigorous space, such as moduli spaces relating fixed string geometries
as well as topological transitions between them.

To that end, we would like to end with a speculation. In algebraic geometry there is a conjecture,
known as Reid's fantasy, that all Calabi-Yau
manifolds (of fixed dimension) are continuously connected by metric deformations and topology
changing transitions. Many expect Reid's fantasy to be true, and if so there is a structural relationship
between all Calabi-Yau manifolds that relates them to one another. In that case, it is reasonable to speculate
that machine learning techniques might implicitly utilize the structural relationships to achieve better-than-expected learning.
Perhaps we are seeing the first evidence of it with the results presented in this work.

\vspace{.5cm}
\noindent \textbf{Acknowledgements.}
We thank Ana Ach\'ucarro, Kyle Cranmer, Mehmet Demirtas, 
Mohamed El Amine Seddik,  Tej Kanwar, Sven Krippendorf, Andre Lukas, Liam McAllister, Fabian Ruehle, Gary Shiu, Alexander Westphal, and especially Danilo
Rezende for discussions regarding this work. Portions of this work were completed at the Aspen
Center for Physics, which is supported by National Science Foundation grant PHY-1607611. J.H. and C.L. are supported
by NSF CAREER grant
PHY-1848089.

\appendix

\begin{figure*}[t]
\begin{center}
\includegraphics[width=.95\columnwidth]{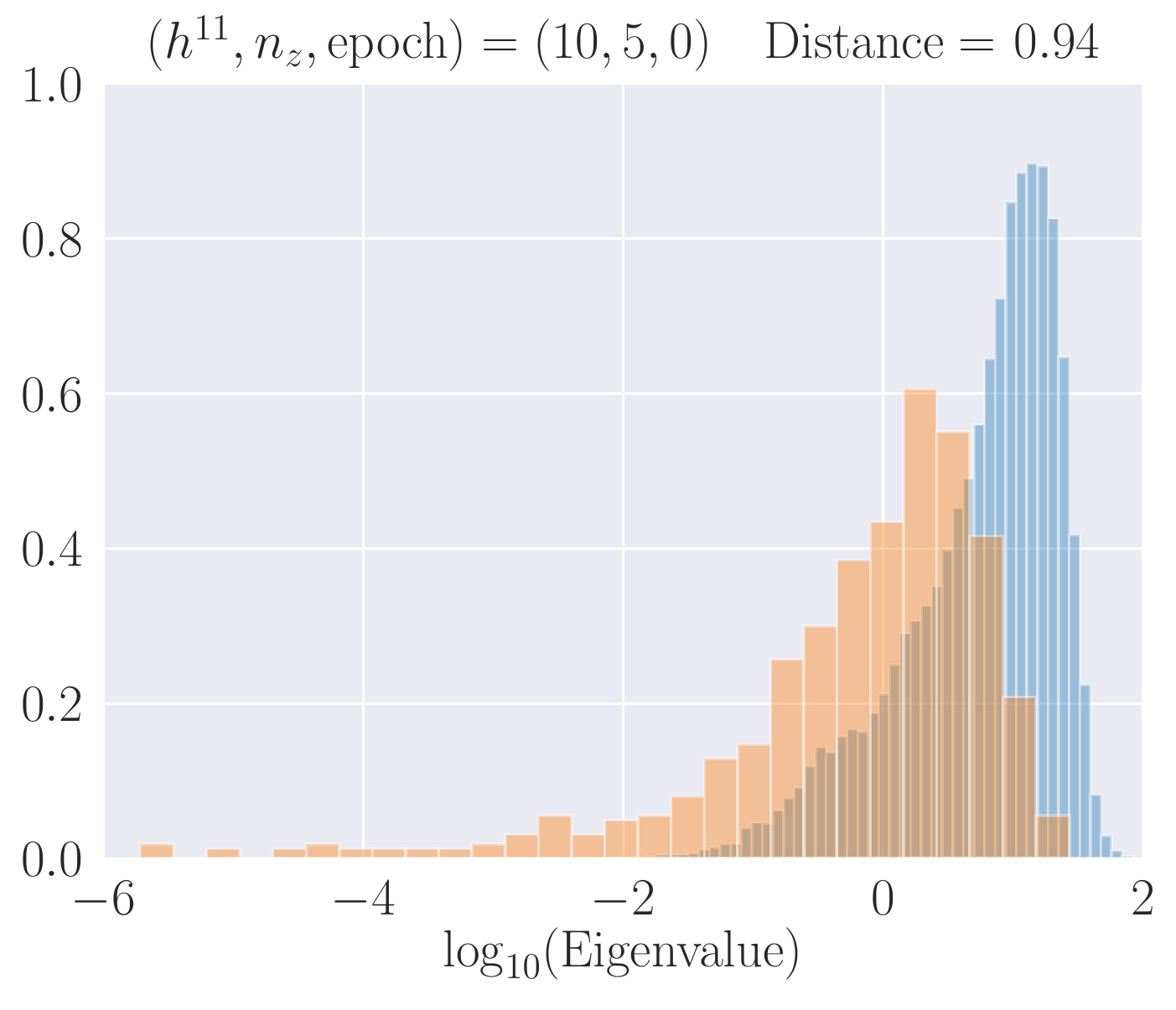} 
\hspace{.5cm} \includegraphics[width=.95\columnwidth]{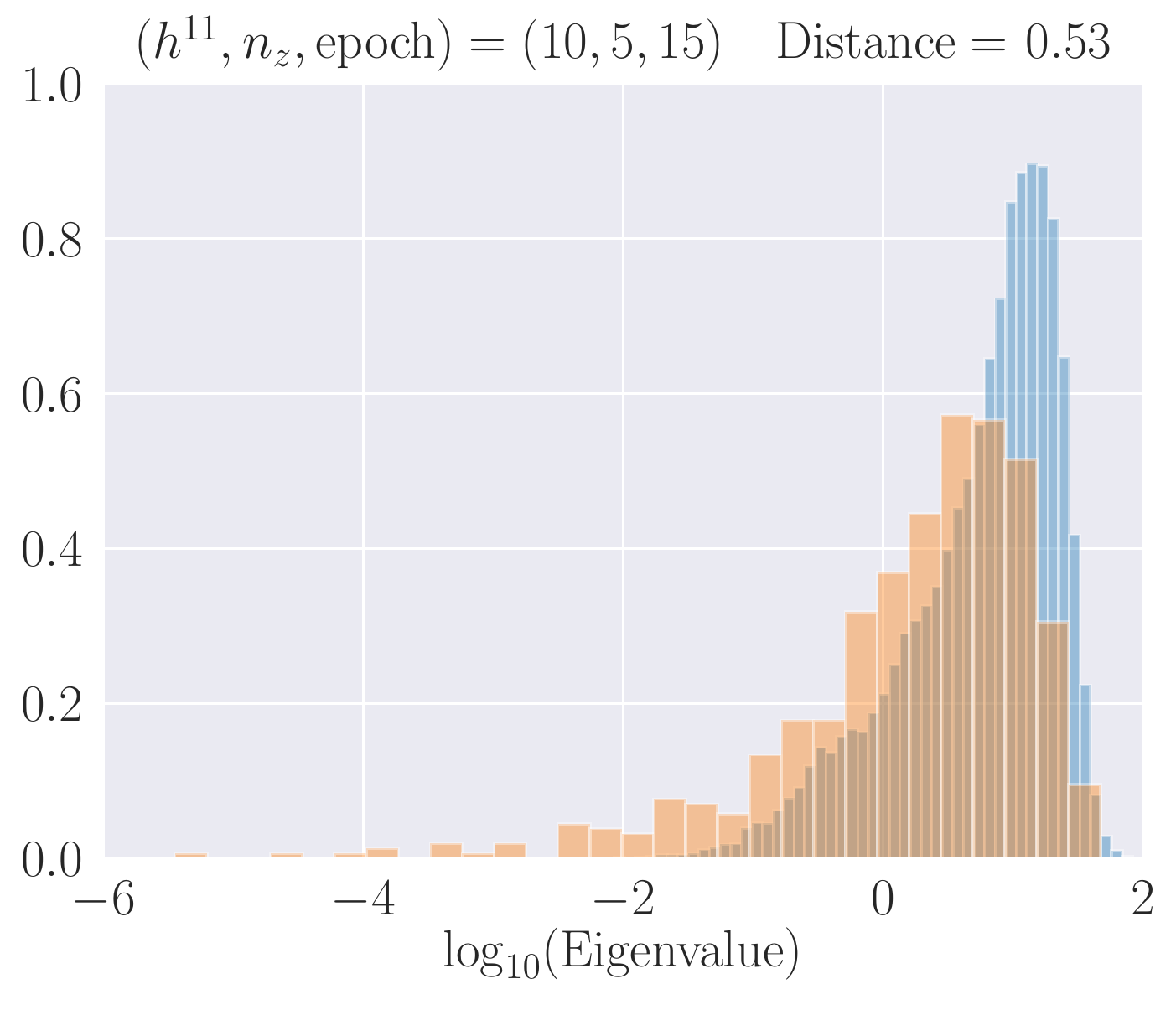} 
\\
\includegraphics[width=.95\columnwidth]{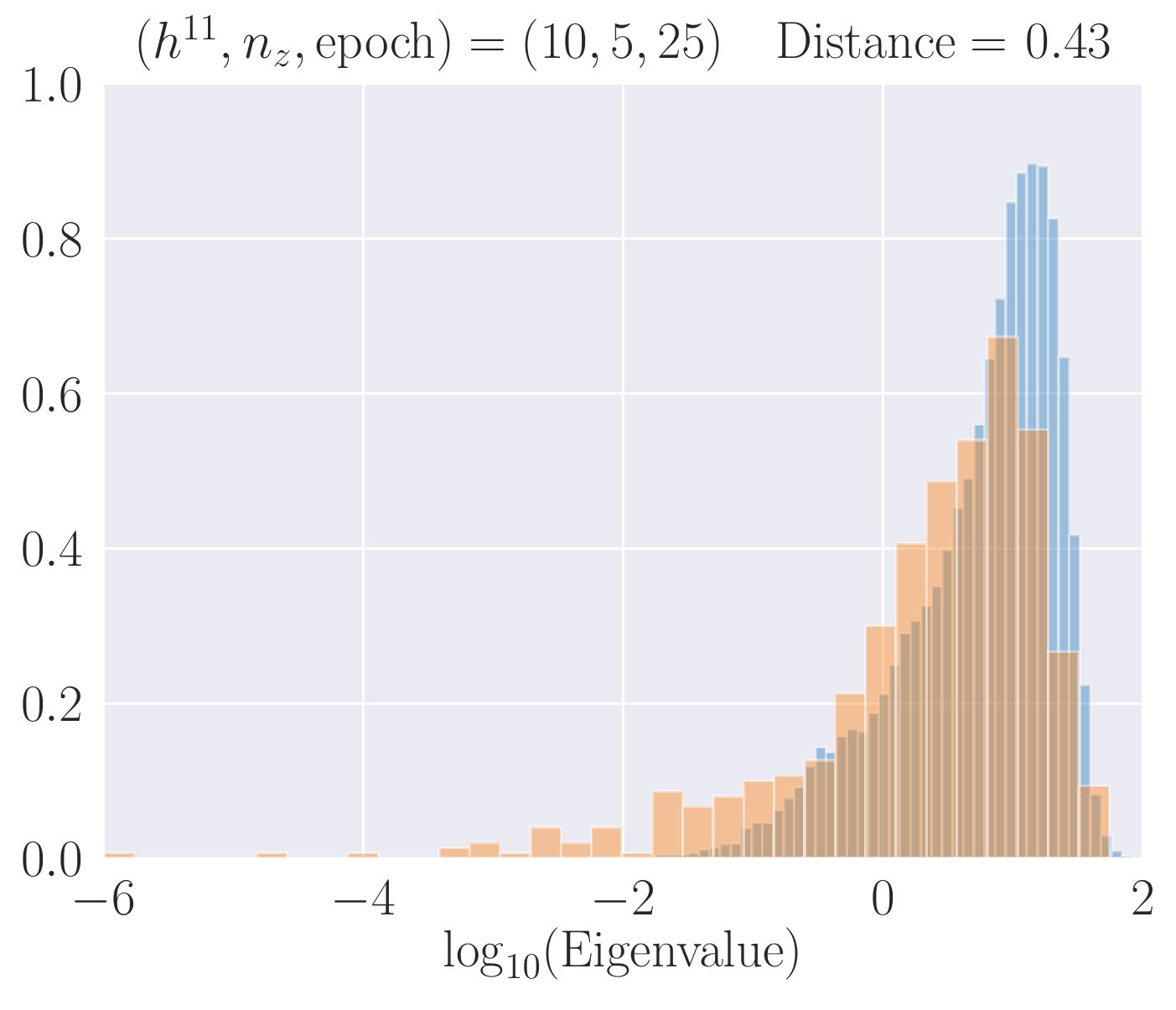} 
\hspace{.5cm}
\includegraphics[width=.95\columnwidth]{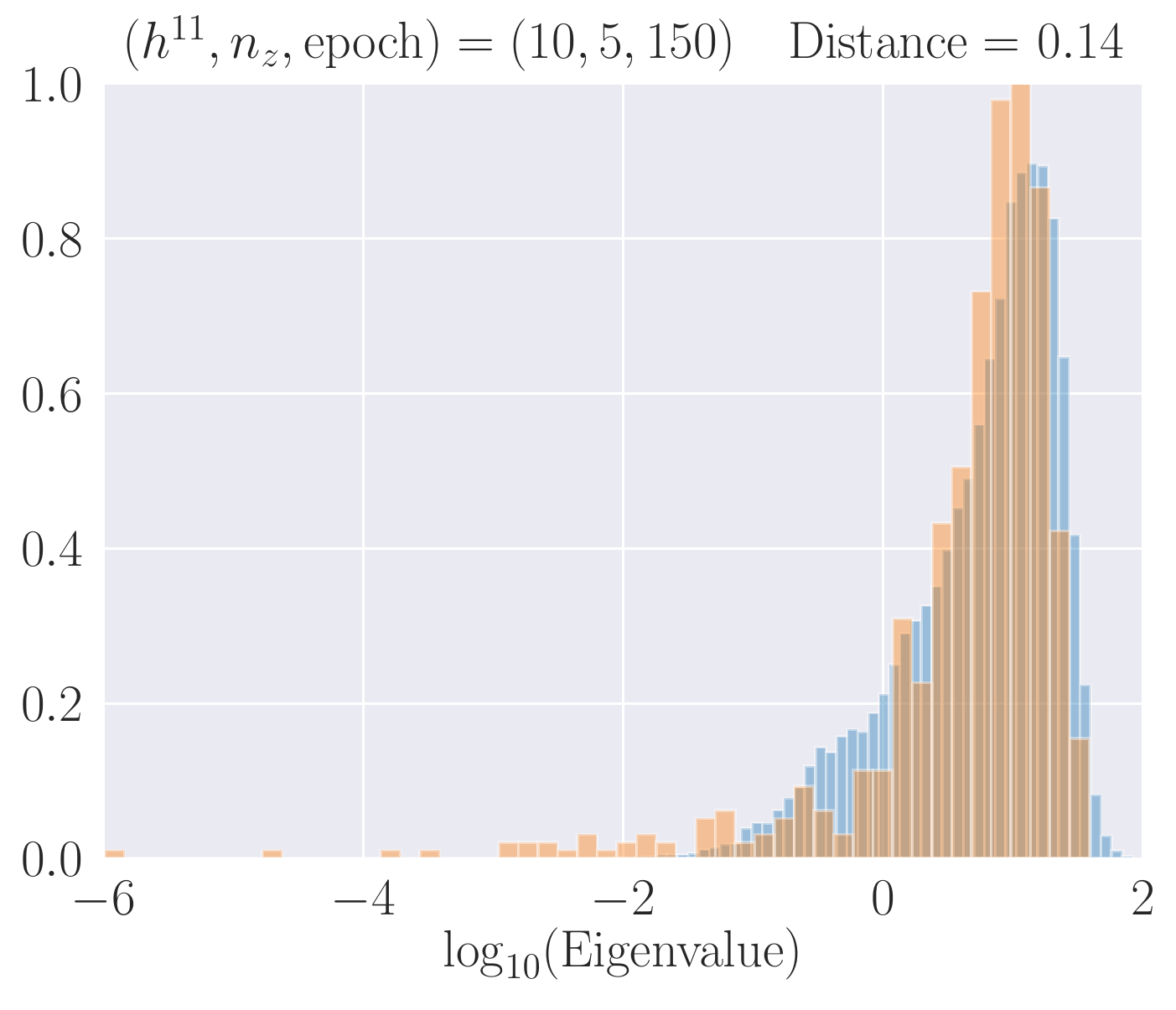} \\ 
\includegraphics[width=.95\columnwidth]{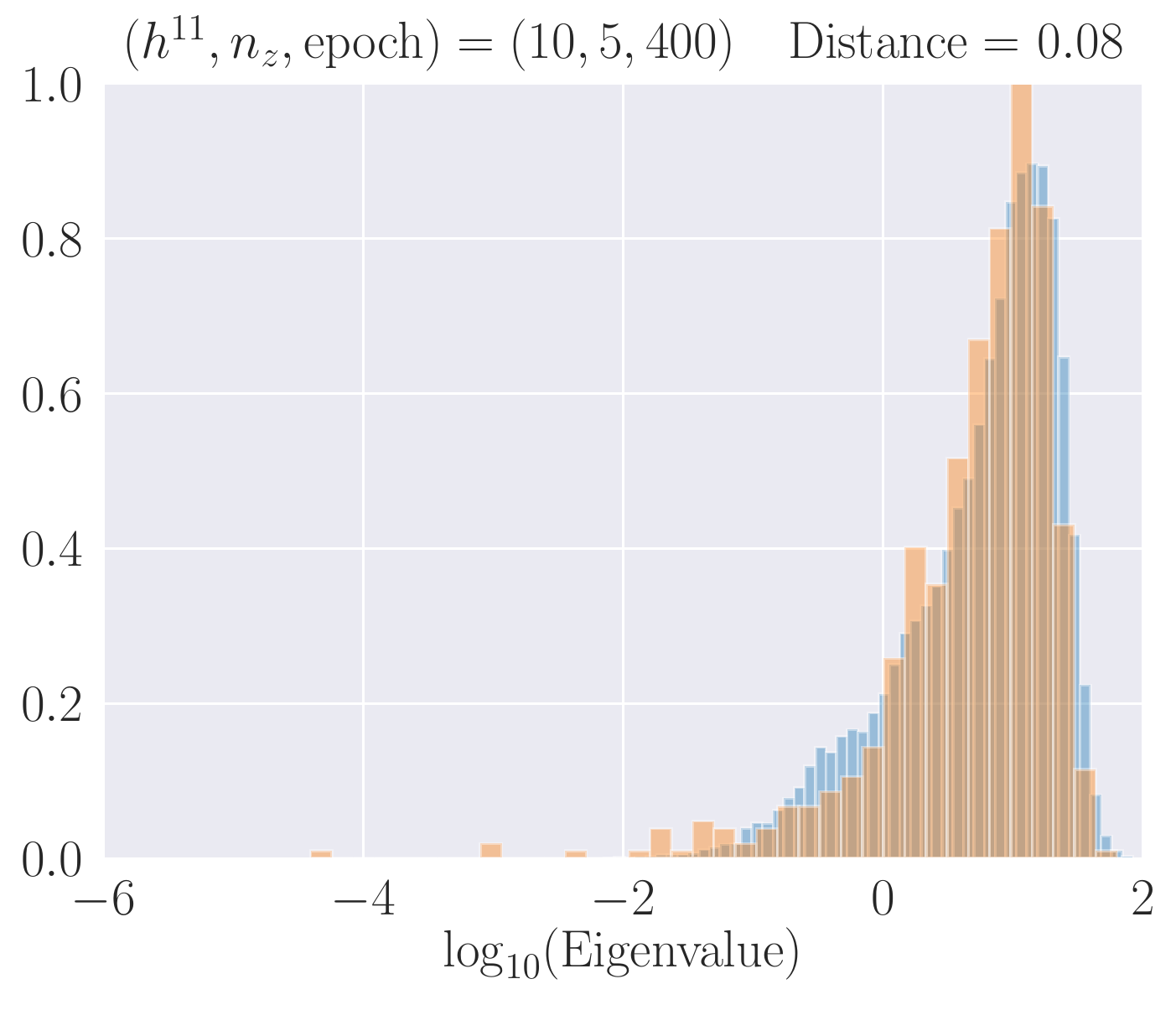} 
\hspace{.5cm}
\includegraphics[width=.95\columnwidth]{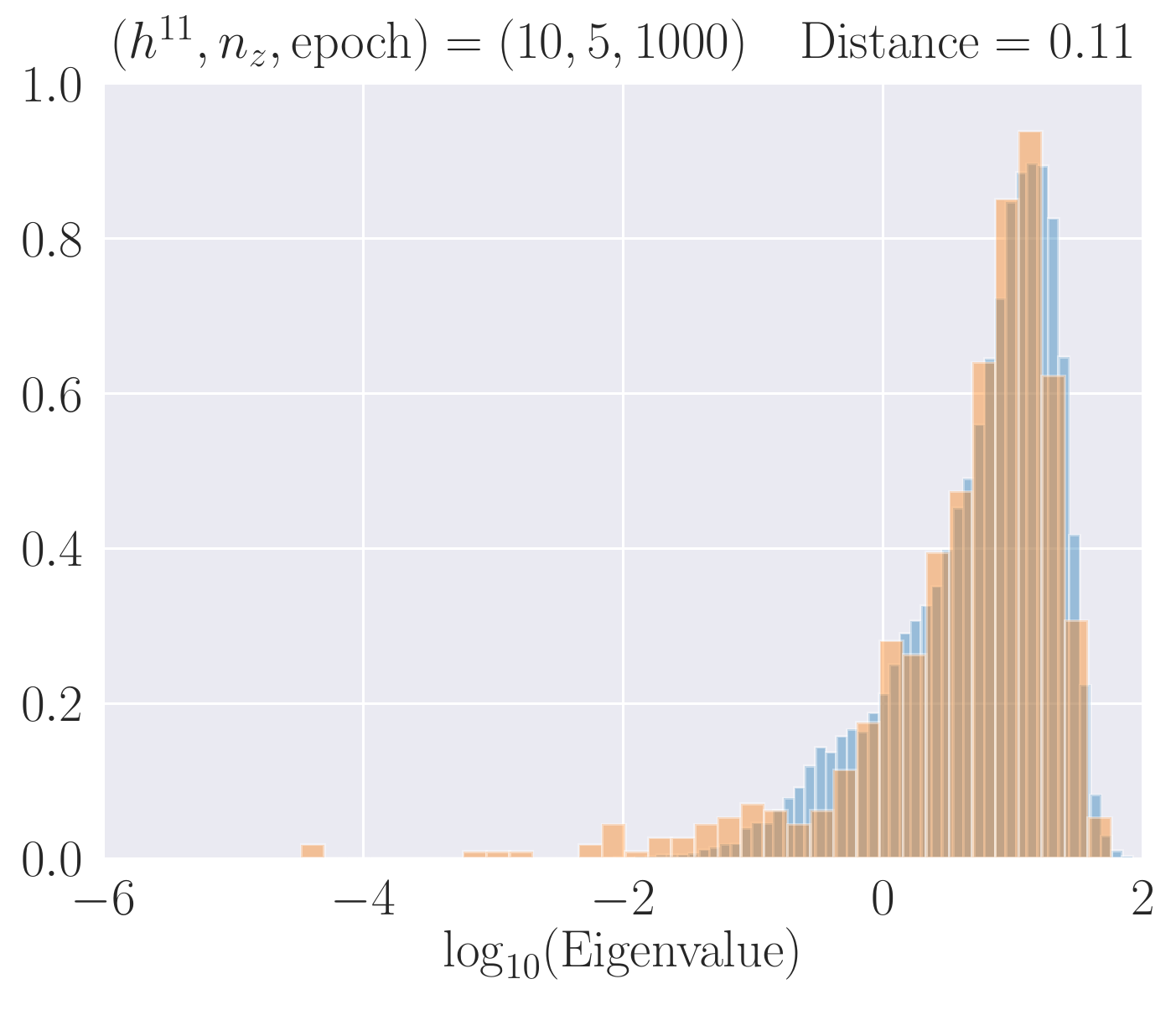} 
\end{center}
\caption{Eigenspectrum change under training an $n_z=5$ Wasserstein DCGAN with $h^{11}=10$.
Blue is the ground truth
K\" ahler metric eigenspectrum from Calabi-Yau compactification.
Orange is the eigenspectrum of the simulated K\" ahler metrics. 
Note the overshoot in the peak while approaching epoch $400$, but its subsequent
flattening as training continues to epoch $1000$. } 
\label{fig:spectrum_change}
\end{figure*}

\begin{figure*}[t]
\centering
\includegraphics[width=.8\columnwidth]{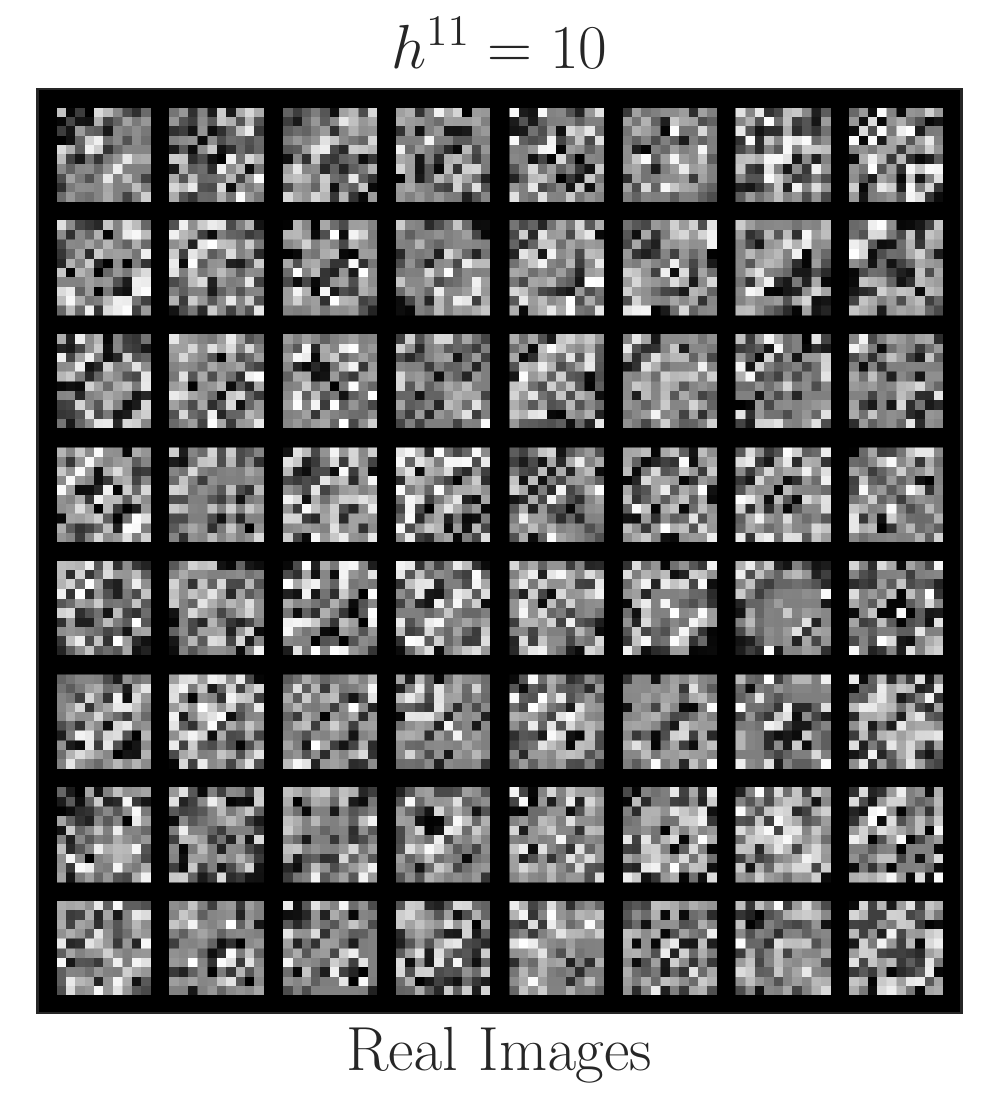}
\hspace{.5cm}
\includegraphics[width=.8\columnwidth]{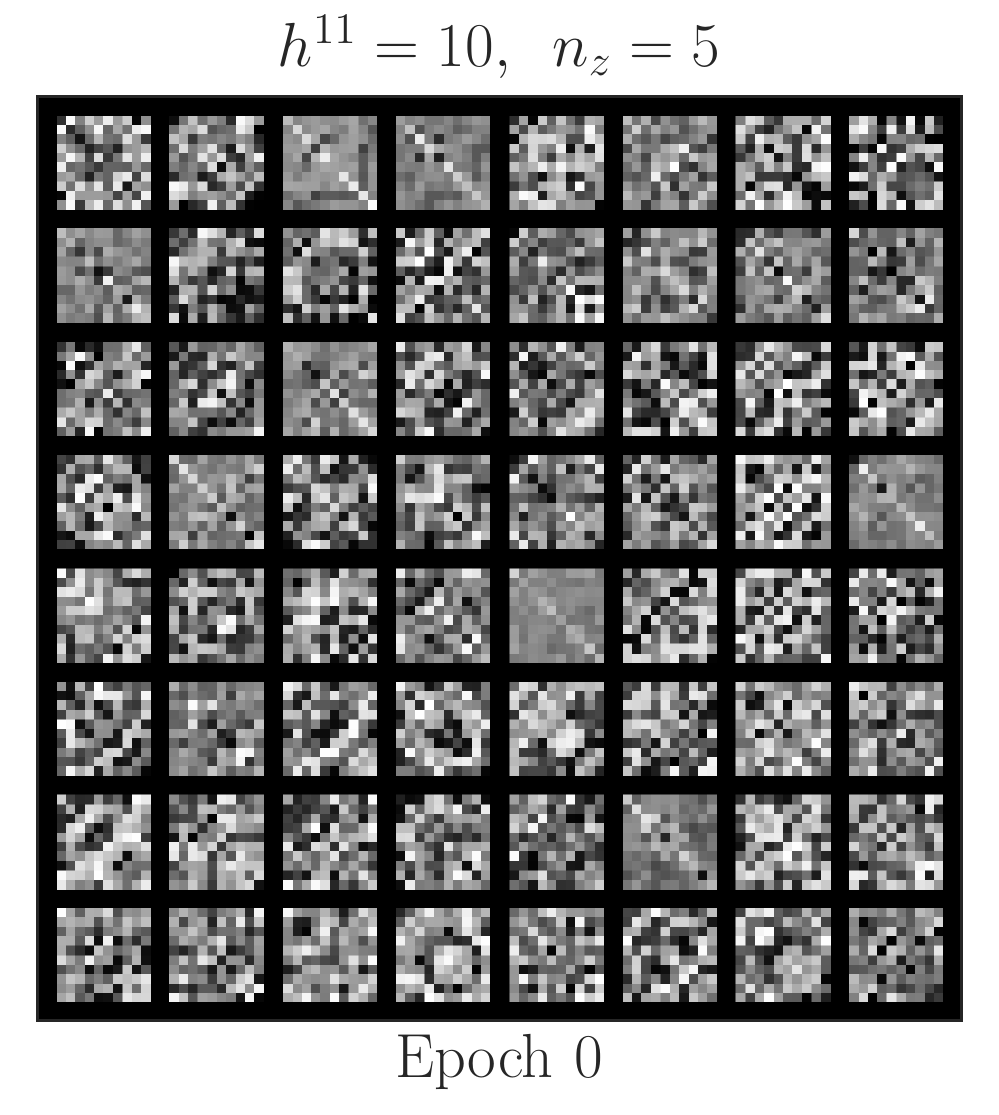} \\
\includegraphics[width=.8\columnwidth]{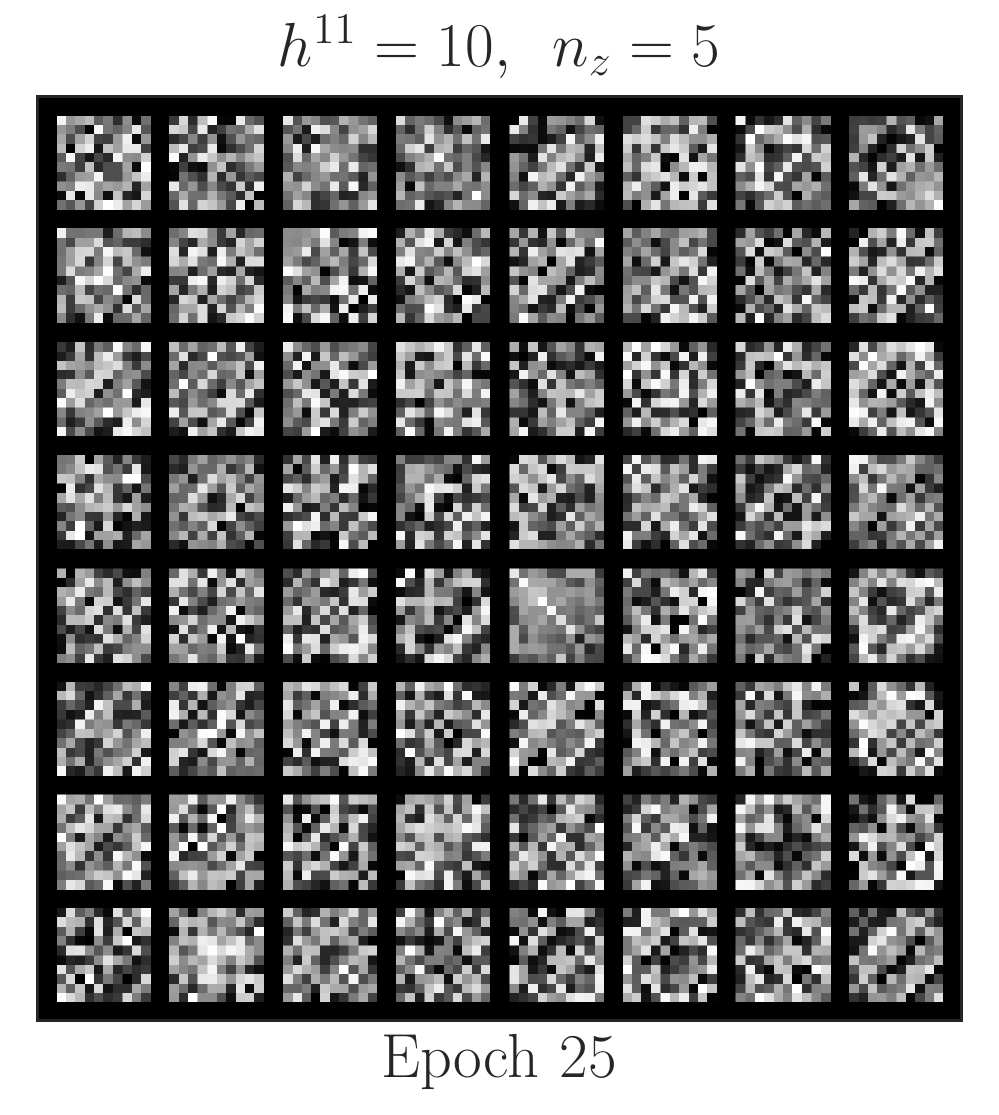} 
\hspace{.5cm}
\includegraphics[width=.8\columnwidth]{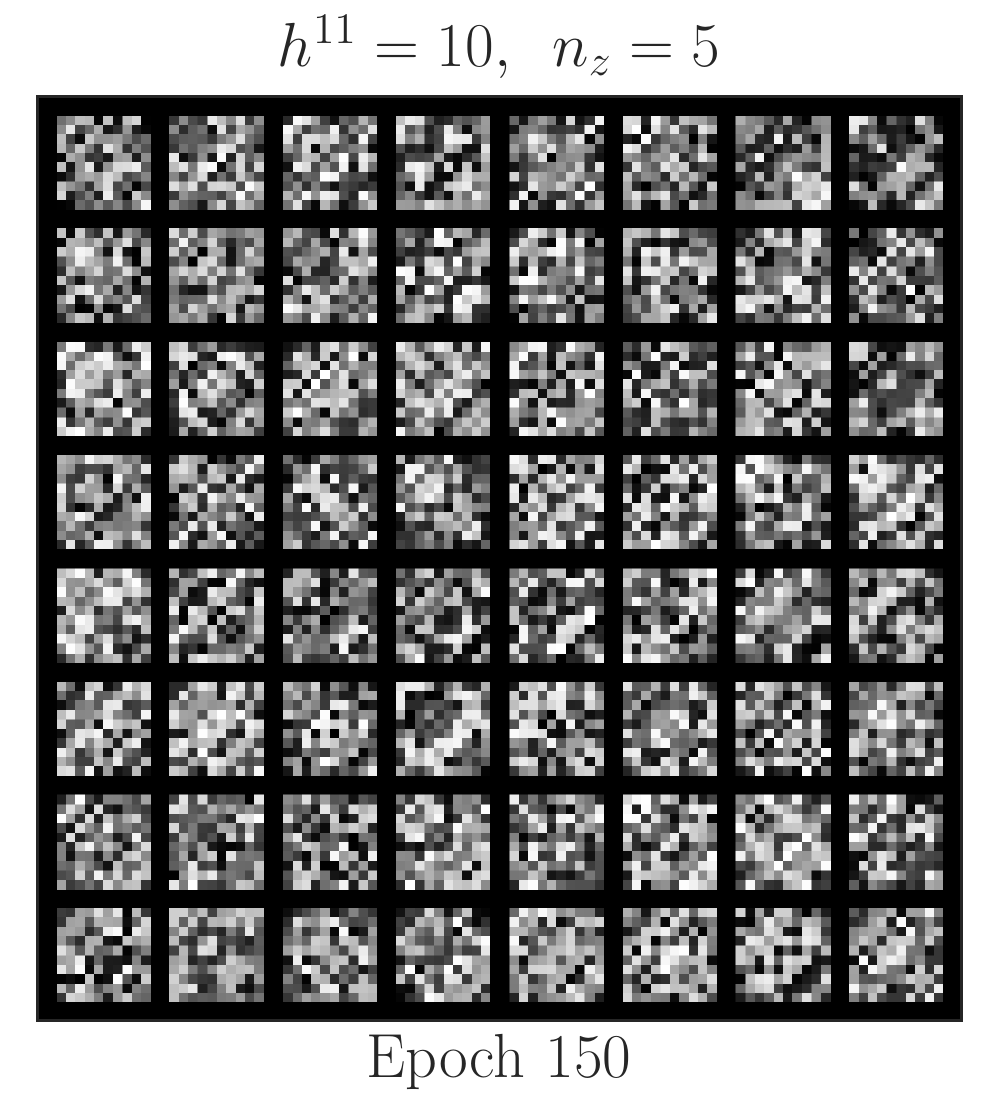} 
\includegraphics[width=.8\columnwidth]{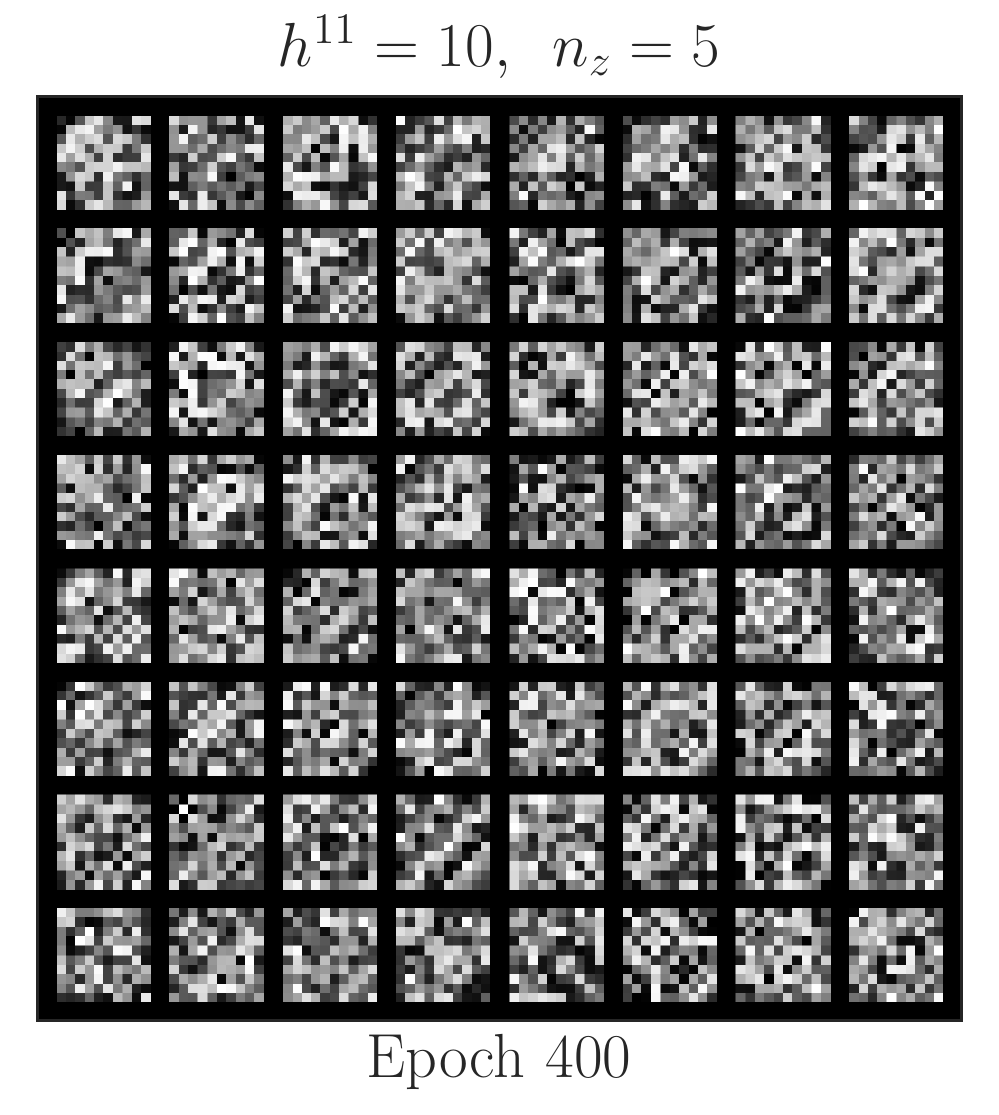} 
\hspace{.5cm}
\includegraphics[width=.8\columnwidth]{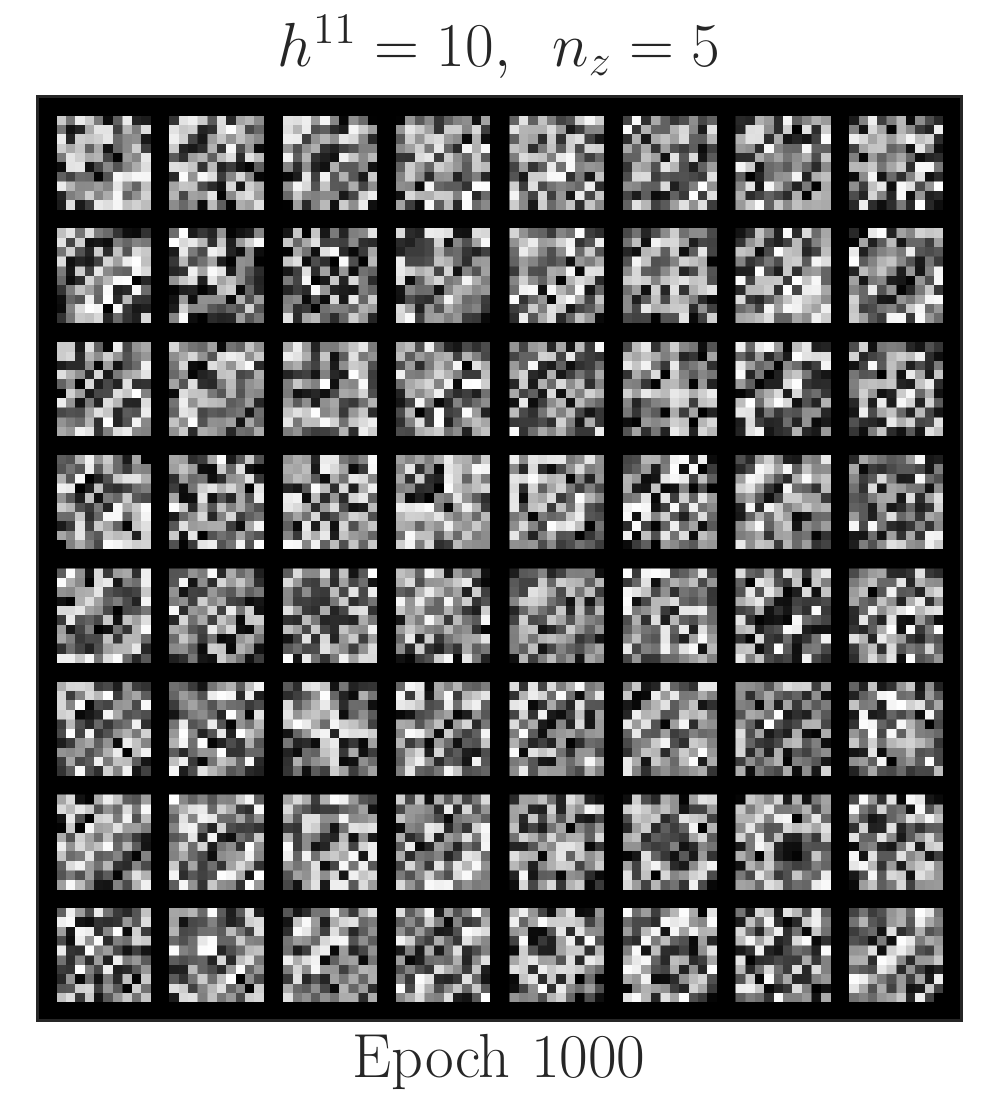} 
\caption{Image representation with fixed noise inputs under training an $n_z=5$ Wasserstein DCGAN with $h^{11}=10$, with
ground truth K\" ahler metrics in the upper left and the rest simulation. Each graphic presents
$64$ K\" ahler metrics, each a $10\times 10$ image. Samples that are faint
and blurry at early times become increasingly sharp and realistic during training.} 
\label{fig:image_change}
\end{figure*}

\bibliography{refs}

\end{document}